\documentclass[twocolumn,twocolappendix,trackchanges]{aastex701}
\usepackage{amsmath}	
\usepackage{bm}
\usepackage{xeCJK}      
\newcommand{\anote}{^{\mathrm{a}}}
\newcommand{\bnote}{^{\mathrm{b}}}
\newcommand{\cnote}{^{\mathrm{c}}}
\newcommand{\dnote}{^{\mathrm{d}}}
\newcommand{\enote}{^{\mathrm{e}}}
\newcommand{\fnote}{^{\mathrm{f}}}
\newcommand{\gnote}{^{\mathrm{g}}}
\newcommand{\hnote}{^{\mathrm{h}}}
\newcommand{\inote}{^{\mathrm{i}}}
\newcommand{\jnote}{^{\mathrm{j}}}
\newcommand{\knote}{^{\mathrm{k}}}
\newcommand{\lnote}{^{\mathrm{l}}}
\newcommand{\mnote}{^{\mathrm{m}}}
\newcommand{\nnote}{^{\mathrm{n}}}
\newcommand{\onote}{^{\mathrm{o}}}
\newcommand{\pnote}{^{\mathrm{p}}}
\newcommand{\qnote}{^{\mathrm{q}}}

\newcommand{\avgrhod}{\overline{\rhod}}                      
\newcommand{\cs}{c_\mathrm{s}}                               
\newcommand{\pareff}{\varepsilon_\parallel}                  
\newcommand{\relpareff}{\varepsilon_\parallel^\mathrm{rel}}  
\newcommand{\etavK}{\eta v_\mathrm{K}}                       
\newcommand{\kWh}{\mathrm{kW}\cdotp\mathrm{h}}               
\newcommand{\meanavgrhod}{\overline{\avgrhod}}               
\newcommand{\np}{n_\mathrm{p}}                               
\newcommand{\od}{\mathrm{d}}                                 
\newcommand{\Prob}{\mathrm{P}}                               
\newcommand{\Npe}{N_\mathrm{PE}}                             
\newcommand{\rhod}{\rho_\mathrm{d}}                          
\newcommand{\rhog}{\rho_\mathrm{g}}                          
\newcommand{\rhogn}{\rho_\mathrm{g,0}}                       
\newcommand{\SMA}{\mathrm{SMA}}                              
\newcommand{\taus}{\tau_\mathrm{s}}                          
\newcommand{\tlim}{t_\mathrm{lim}}                           
\newcommand{\TNpe}{T_{\Npe}}                                 
\newcommand{\tstop}{t_\mathrm{stop}}                         
\newcommand{\Uvec}{\mathbf{U}}                               
\newcommand{\uvec}{\mathbf{u}}                               
\newcommand{\Vvec}{\mathbf{V}}                               
\newcommand{\vvec}{\mathbf{v}}                               
\newcommand{\xhat}{\hat{\mathbf{x}}}                         
\newcommand{\yhat}{\hat{\mathbf{y}}}                         
\newcommand{\avgE}{\overline{E}}                             
\newcommand{\dhrhod}{\hat{\delta}\rhod}                      
\newcommand{\meanavgE}{\overline{\avgE}}                     
\newcommand{\Eshot}{E_\mathrm{shot}}                         
\newcommand{\kNyq}{k_\mathrm{Nyq}}                           

\begin{document}

\title{A Comparative Study of the Streaming Instability:\\Unstratified Models with Marginally Coupled Grains}

\correspondingauthor{Stanley A. Baronett}
\author[orcid=0000-0003-0412-760X,gname=Stanley,sname=Baronett]{Stanley A. Baronett}
\altaffiliation{VITA--VICO Fellow}
\affiliation{Department of Astronomy, University of Virginia, Charlottesville, VA 22904, USA}
\email[show]{vvu2tz@virginia.edu}

\author[orcid=0000-0002-3768-7542,gname=Wladimir,sname=Lyra]{Wladimir Lyra}
\affiliation{Department of Astronomy, New Mexico State University, Las Cruces, NM 88003, USA}
\email{wlyra@nmsu.edu}

\author[orcid=0000-0002-1342-1694,gname=Hossam,sname=Aly]{Hossam Aly}
\affiliation{Faculty of Aerospace Engineering, Delft University of Technology, Delft, 2629 HS, The Netherlands}
\email{hossam.saed@gmail.com}

\author[orcid=0009-0006-2478-5246,gname=Olivia,sname=Brouillette]{Olivia Brouillette}
\affiliation{Department of Astronomy, New Mexico State University, Las Cruces, NM 88003, USA}
\email{ob1@nmsu.edu}

\author[orcid=0000-0001-6259-3575,gname=Daniel,sname=Carrera]{Daniel Carrera}
\affiliation{Department of Astronomy, New Mexico State University, Las Cruces, NM 88003, USA}
\email{carrera4@nmsu.edu}

\author[gname=Victoria,sname=De Cun]{Victoria I. De Cun}
\affiliation{Department of Astronomy, New Mexico State University, Las Cruces, NM 88003, USA}
\email{videcun@gmail.com}

\author[orcid=0000-0002-8247-6453,gname=Linn,sname=Erikssonn]{Linn E. J. Eriksson}
\affiliation{Department of Astrophysics, American Museum of Natural History, New York, NY 10024, USA}
\email{leriksson@amnh.org}

\author[orcid=0000-0002-9298-3029,gname=Mario,sname=Flock]{Mario Flock}
\affiliation{Max‑Planck‑Institut für Astronomie, Heidelberg, 69117, Germany}
\email{flock@mpia.de}

\author[orcid=0000-0002-7575-3176,gname=Pinghui,sname=Huang]{Pinghui Huang (黄平辉)}
\affiliation{CAS Key Laboratory of Planetary Sciences, Purple Mountain Observatory, Chinese Academy of Sciences, Nanjing, 210023, China}
\email{phhuang@pmo.ac.cn}

\author[orcid=0000-0001-7671-9992,gname=Leonardo,sname=Krapp]{Leonardo Krapp}
\affiliation{Departamento de Astronomía, Universidad de Concepción, Concepción, 4030000, Chile}
\email{lkrapp@udec.cl}

\author[orcid=0000-0002-8896-9435,gname=Geoffroy,sname=Lesur]{Geoffroy Lesur}
\affiliation{Université Grenoble Alpes, CNRS, IPAG, Grenoble, 38058, France}
\email{geoffroy.lesur@univ-grenoble-alpes.fr}

\author[orcid=0000-0001-9222-4367,gname=Rixin,sname=Li]{Rixin Li (李日新)}
\affiliation{Department of Astronomy, Theoretical Astrophysics Center, and Center for Integrative Planetary Science, University of California, Berkeley, Berkeley, CA 94720, USA}
\email{rixin@berkeley.edu}

\author[orcid=0000-0002-4142-3080,gname=Shengtai,sname=Li]{Shengtai Li (李胜台)}
\affiliation{Los Alamos National Laboratory, Los Alamos, NM 87545, USA}
\email{sli@lanl.gov}

\author[orcid=0000-0003-2719-6640,gname=Jeonghoon,sname=Lim]{Jeonghoon Lim (임정훈)}
\affiliation{Nevada Center for Astrophysics and Department of Physics and Astronomy, University of Nevada, Las Vegas, Las Vegas, NV 89154, USA}
\email{jay.lim1@unlv.edu}

\author[orcid=0000-0002-8378-7608,gname=Sijme-Jan,sname=Paardekooper]{Sijme-Jan Paardekooper}
\affiliation{Faculty of Aerospace Engineering, Delft University of Technology, Delft, 2629 HS, The Netherlands}
\email{s.paardekooper@tudelft.nl}

\author[orcid=0000-0002-5000-2747,gname=David,sname=Rea]{David G. Rea}
\affiliation{Department of Physics and Astronomy, Iowa State University, Ames, IA 50011, USA}
\email{drea1@iastate.edu}

\author[orcid=0000-0003-0801-3159,gname=Debanjan,sname=Sengupta]{Debanjan Sengupta}
\affiliation{Department of Astronomy, New Mexico State University, Las Cruces, NM 88003, USA}
\email{debanjan@nmsu.edu}

\author[orcid=0000-0002-3771-8054,gname=Jacob,sname=Simon]{Jacob B. Simon}
\affiliation{Department of Physics and Astronomy, Iowa State University, Ames, IA 50011, USA}
\email{jbsimon@iastate.edu}

\author[orcid=0000-0001-9073-497X,gname=Prakruti,sname=Sudarshan]{Prakruti Sudarshan}
\affiliation{Max‑Planck‑Institut für Astronomie, Heidelberg, 69117, Germany}
\email{sudarshan@mpia.de}

\author[orcid=0000-0001-5372-4254,gname=Orkan,sname=Umurhan]{Orkan M. Umurhan}
\affiliation{SETI Institute, Mountain View, CA 94043, USA}
\email{oumurhan@seti.org}

\author[orcid=0000-0003-2589-5034,gname=Chao-Chin,sname=Yang]{Chao-Chin Yang (楊朝欽)}
\affiliation{Department of Physics and Astronomy, The University of Alabama, Tuscaloosa, AL 35487, USA}
\email{ccyang@ua.edu}

\author[orcid=0000-0002-3644-8726,gname=Andrew,sname=Youdin]{Andrew N. Youdin}
\affiliation{Lunar and Planetary Laboratory and Department of Astronomy and Steward Observatory, University of Arizona, Tucson, AZ 85721, USA}
\email{youdin@arizona.edu}


\begin{abstract}

The streaming instability is a leading mechanism for concentrating solids and initiating planetesimal formation in protoplanetary disks.
Although numerous studies have explored its linear growth, nonlinear evolution, and implications for planet formation, the diversity of numerical methods and dust treatments used across the literature has made it difficult to assess which features of the instability are physically robust and which arise from code-dependent choices.
We present the first systematic comparison of seven hydrodynamic codes---spanning finite-volume and finite-difference schemes and modeling dust either as Lagrangian particles or as a pressureless fluid---applied to the unstratified streaming instability with a dimensionless stopping time of unity.
All codes reproduce the characteristic sequence of exponential growth, filament formation, and turbulent saturation, demonstrating broad agreement in the qualitative behavior of the instability.
Quantitatively, however, the dust model remains the dominant source of variation at moderate resolution:
particle-based simulations reach higher peak densities and exhibit broader high-density tails than fluid-based models at $512^2$ resolution, although increasing the number of particles brings their initial maximum density evolution into close agreement with that of dust-fluid models.
At $1024^2$, these differences diminish substantially, indicating better agreement of the saturated-state statistics across dust treatments.
In terms of computational performance, most particle implementations suffer from imbalanced parallelized loads, while execution on a GPU is at least two to three times more energy efficient and scales better at higher resolutions than on CPUs.
Given the intrinsic stochasticity of this nonlinear system, only statistical diagnostics remain meaningful across codes.

\end{abstract}



\section{Introduction} \label{sec:introduction}

The streaming instability is regarded as a key aerodynamic mechanism for facilitating planetesimal formation within the midplane dust layer of protoplanetary disks.
\cite{YoudinGoodman2005} first identified it as an unstratified linear instability arising from momentum exchange between gas and monodisperse dust, and subsequent numerical investigations have explored its parameter dependence \citep{Pencil_dust} and nonlinear evolution \citep{JohansenYoudin2007,BaronettYangZhu2024,LimBaronettSimon2025}.
Although the original analysis assumed incompressible hydrodynamics, \cite{Pencil_dust} extended their linear analysis to include compressibility,
and most simulations---beginning with \cite{JohansenYoudin2007}---have since adopted compressible gas dynamics, including those presented here.

Several works have sought to clarify the physical mechanism underlying this instability, with an initial effort predating its formal characterization.
\cite{GoodmanPindor2000} provided a foundational overview of drag-feedback instabilities, whereas \cite{LinYoudin2017} refined the analysis by incorporating thermodynamic work (i.e., compressional heating).
Subsequent studies indicated that the mechanism described by \cite{SquireHopkins2018a} exhibits similarities to a broader set of drag-mediated instabilities that arise when solids drift through gas \citep{SquireHopkins2018b,HopkinsSquire2018}.\footnote{
In the general resonant-drag-instability formulation, growth rates formally diverge as the solid-to-gas ratio approaches or exceeds unity, so the applicability of the framework in that regime is uncertain.}
More recently, \cite{MagnanHeinemannLatter2024} examined the low dust-to-gas density–ratio regime in detail, demonstrating that the streaming instability remains active even when solids are extremely dilute and clarifying the limiting linear behavior of the system in this regime.
Despite these advances, the nonlinear saturation phase of the instability---particularly the structure, evolution, and merging of dust filaments and the resulting saturated density distribution---is not completely understood, with several studies showing that precise characterization of these features requires high-resolution numerical experiments (e.g., \citealt{JohansenYoudin2007}; \citealt{Athena_particles}; \citealt{YangJohansen2014}; \citealt{BaronettYangZhu2024}).
Compounding this challenge, methodological discrepancies---such as differences between finite-difference and finite-volume schemes, and between dust treatments as pressureless fluids versus Lagrangian particles---introduce varying results, making it difficult to disentangle genuine physical effects from numerical artifacts.

In an effort to address this difficulty, we present the first systematic comparison of hydrodynamic codes that model the nonlinear evolution of the streaming instability.
Section~\ref{sec:methodology} outlines the governing equations for gas and dust in an unstratified disk model, the problem setup, and the numerical methods implemented across the selected codes.
In Section~\ref{sec:results}, we examine the morphologies, temporal evolution, and saturated states produced by each code, as well as their computational performance.
Section~\ref{sec:discussion} considers the role of turbulence in the nonlinear regime and discusses the implications of our findings for dust modeling approaches.
Finally, Section~\ref{sec:conclusions} summarizes the key findings and highlights directions for future work.

\section{Methodology} \label{sec:methodology}

We use seven different hydrodynamic codes to solve the nonlinear evolution of the streaming instability and its turbulent saturated state.
Section~\ref{sec:unstratified_disk_model} specifies the unstratified model and the problem of interest investigated previously.
Section~\ref{sec:codes_numerical_methods} discriminates the codes and numerical methods used in this comparison.

\subsection{Unstratified Disk Model} \label{sec:unstratified_disk_model}

The model uses the local-shearing-box approximation \citep{GoldreichLynden-Bell1965} for a system of gas and dust within a non-magnetized region of a protoplanetary disk.
The scale-invariant, axisymmetric computational domain orbits about the orbital angular vector $\bm{\Omega}$ at the local Keplerian frequency $\Omega$ at an arbitrary radial distance $r_0$ from the central star.
Assuming the domain size is much smaller than its orbital distance, the equations of motion can be linearized with Cartesian $x$, $y$, and $z$ axes constantly aligned to the radial, azimuthal, and vertical directions, respectively, where $\bm{\Omega} = \Omega\hat{\mathbf{z}}$.

The unstratified model may represent the sedimented dust layer near the midplane of the disk \citep{LimBaronettSimon2025} and omits the vertical component of stellar gravity.
Sections~\ref{sec:gas}, \ref{sec:lagrangian_dust_particles}, and \ref{sec:pressureless_dust_fluid} detail the governing equations for the gas, the dust as Lagrangian particles, and the dust as a pressureless fluid, respectively.
Section~\ref{sec:problem_ba} details the computational setup and parameters we focus on in this work.

\subsubsection{Gas} \label{sec:gas}

The continuity and momentum equations for the inviscid gas, with density $\rhog$ and total velocity $\Uvec$, are 
(cf. \citealt[eqs.~1a and 1c]{HawleyGammieBalbus1995}, and \citealt[eqs.~1 and 2]{LimSimonLi2024})
\begin{align}
    \frac{\partial\rhog}{\partial t} + \nabla\cdot(\rhog\Uvec) = &\;0,
    \label{eq:gas_cont}\\
    \frac{\partial\rhog\Uvec}{\partial t} + \nabla\cdot(\rhog\Uvec\Uvec + P\mathbf{I}) = &\;\rhog\left[3\Omega^2x\xhat + 2\Uvec\times\bm{\Omega}\right.\nonumber\\
    &\left.+\,2\Omega\Pi \cs\xhat - \frac{\rhod}{\rhog}\left(\frac{\Uvec - \Vvec}{\tstop}\right)\right]
    \label{eq:gas_mom}
\end{align}
respectively.
The pressure $P=\cs^2\rhog$ for an isothermal equation of state with sound speed $\cs$, and $\mathbf{I}$ is the identity matrix.
On the right-hand side of equation~\eqref{eq:gas_mom}, the first source term is the tidal term for a Keplerian disk.
The second term is the Coriolis force.
The third is a constant outward force on the gas due to an external radial pressure gradient, set by the dimensionless parameter \citep[cf.][eq.~1]{BaiStone2010}
\begin{equation}
    \Pi\equiv\frac{\etavK}{\cs}=\frac{\eta r}{H},
    \label{eq:Pi}
\end{equation}
where $H = \cs/\Omega$ is the vertical gas scale height, and
\begin{equation}
    \eta \equiv -\frac{1}{2}\frac{1}{\rhog\Omega^2r}\frac{\partial P}{\partial r} = -\frac{1}{2}\left(\frac{H}{r}\right)^2\frac{\partial\ln P}{\partial\ln r} \sim \left(\frac{\cs}{v_\mathrm{K}}\right)^2,
    \label{eq:eta}
\end{equation}
is the fractional reduction (when $\eta > 0$) in $U_y$ from the local Keplerian velocity $v_\mathrm{K} \equiv \Omega r_0$ in the absence of dust, i.e., $U_y = v_\mathrm{K}(1 - \eta)$ \citep[cf.][eq.~1.9]{NakagawaSekiyaHayashi1986}.
The fourth and final term is the mutual drag force between the dust and gas, where $\Vvec$ is the local (ensemble-averaged) dust total velocity, and $\tstop$ is the $e$-folding stopping time to damp the relative speed between a solid particle and the surrounding gas due to friction \citep{Whipple1972, Weidenschilling1977MNRAS}.
The factor of the local dust-to-gas density ratio $\rhod / \rhog$ is required for the conservation of the total linear momentum.

Alternatively, the momentum equation~\eqref{eq:gas_mom} can be recast in terms of the residual velocity $\uvec$.
Splitting the total velocity into
\begin{equation}
    \Uvec = \uvec_0 + \uvec,
    \label{eq:tot_vel}
\end{equation}
with linearized background Keplerian shear flow
\begin{equation}
    \uvec_0 = -(3/2)\Omega x\yhat,
    \label{eq:lin_kep_gas_vel}
\end{equation}
and substituting equation~\eqref{eq:tot_vel} into equation~\eqref{eq:gas_mom} ultimately yields
(cf. \citealt[eq.~4]{Pencil_dust}, and \citealt[eq.~2]{BaronettYangZhu2024})
\begin{align}
    \frac{\partial\rhog\uvec}{\partial t} + \nabla\cdot(\rhog\uvec\uvec + P\mathbf{I}) = &\;\rhog\left[2\Omega u_y\xhat- \frac{1}{2}\Omega u_x\yhat\right.\nonumber\\
    &\left.+\,2\Omega\Pi \cs\xhat - \frac{\rhod}{\rhog}\left(\frac{\uvec - \vvec}{\tstop}\right)\right],
    \label{eq:gas_mom_resid}
\end{align}
where $\vvec$ is the local (ensemble-averaged) dust residual velocity.
Due to axisymmetry, the advection term by the Keplerian shear, $u_0\partial/\partial y$, is not present in the left-hand side of equation~\eqref{eq:gas_mom_resid} \citep{BrandenburgNordlundStein1995}.

The gas density field is initially uniform, such that $\rhog(x,y,z,t=0) = \rhogn$.
By assuming a total dust-to-gas mass ratio
\begin{equation}
    \epsilon \equiv \frac{\langle\rhod\rangle}{\rhogn},
    \label{eq:epsilon}
\end{equation} where
\begin{equation}
    \langle\rhod\rangle \equiv \frac{1}{L_x L_y L_z} \iiint \rhod\od x\od y\od z
    \label{eq:vol_avg}
\end{equation}
is the volume average over the dimensions of the computational domain $L_x\times L_y\times L_z$, the initial components of the gas velocity take the equilibrium solution by \citet{NakagawaSekiyaHayashi1986} \citep[cf.][eqs.~7a-7b]{Pencil_dust}:
\begin{align}
    u_{x,0} &= \frac{2\epsilon\taus}{(1 + \epsilon)^2 + \taus^2}\etavK,
    \label{eq:u_x,0}\\
    u_{y,0} &= -\left[1 + \frac{\epsilon\taus^2}{(1 + \epsilon)^2 + \taus^2}\right] \frac{\etavK}{1 + \epsilon},
    \label{eq:u_y,0}\\
    u_{z,0} &= 0,
    \label{eq:u_z,0}
\end{align}
where
\begin{equation}
    \taus \equiv \Omega \tstop
    \label{eq:taus}
\end{equation}
is the dimensionless stopping time \citep[i.e., Stokes number;][]{YoudinGoodman2005}.

\subsubsection{Lagrangian Dust Particles}
\label{sec:lagrangian_dust_particles}

For dust modeled as Lagrangian super-particles, where each one represents an ensemble of numerous identical solid particles described by their total mass and average total velocity, the equation of motion for the $i$-th particle is \citep[cf.][eq.~4]{LimSimonLi2024}
\begin{equation}
    \frac{\od\Vvec_i}{\od t} = 3\Omega^2x_{\mathrm{p},i}\xhat + 2\Vvec_i \times \bm{\Omega} - \frac{\Vvec_i - \Uvec}{\tstop}.
    \label{eq:par_acc}
\end{equation}
Alternatively, the equations of motion for the dust can be recast in terms of the residual velocity $\vvec_i$, where total velocity $\Vvec_i = \vvec_{i,0} + \vvec_i$, such that \citep[cf.][eqs.~11 and 12]{Pencil_dust}
\begin{align}
    \frac{\od\mathbf{x}_{\mathrm{p},i}}{\od t} &= \vvec_i - \frac{3}{2}\Omega x_{\mathrm{p},i} \yhat,
    \label{eq:par_vel_resid}\\
    \frac{\od\vvec_i}{\od t} &= 2\Omega v_{i,y}\xhat - \frac{1}{2}\Omega v_{i,x}\yhat - \frac{\vvec_i - \uvec}{\tstop},
    \label{eq:par_acc_resid}
\end{align}
are measured relative to the linearized background Keplerian shear flow $\vvec_{i,0} = -(3/2)\Omega x_{\mathrm{p},i}\yhat$ [cf. equation~\eqref{eq:lin_kep_gas_vel}].
The gas velocities $\Uvec$ or $\uvec$ are interpolated at the particle position $\mathbf{x}_{\mathrm{p},i}$.
The right-hand sides of equations~\eqref{eq:par_acc} and \eqref{eq:par_acc_resid} parallel those of \eqref{eq:gas_mom} and \eqref{eq:gas_mom_resid} in Lagrangian form, respectively, without the radial gas pressure gradient.\footnote{Although some codes here (Table~\ref{tab:codes}) instead subtract the background pressure gradient [i.e., the third source term in equation~\eqref{eq:gas_mom}] from the right-hand sides of equations~\eqref{eq:gas_mom} and \eqref{eq:par_acc} or \eqref{eq:dust_mom}, we note that the total effect is mathematically equivalent.\label{foot:pressure_gradient}}
For a monodisperse dust population, the stopping times $\tstop$ and $\taus$ [equation~\eqref{eq:taus}] are the same for all particles.
As with the gas [cf. equations~\eqref{eq:u_x,0}--\eqref{eq:u_z,0}], the initial components of the particle velocity take the equilibrium solution by \citet{NakagawaSekiyaHayashi1986} \citep[cf.][eqs.~7c-7d]{Pencil_dust}:
\begin{align}
    v_{i,x,0} &= -\left[\frac{2\taus}{(1 + \epsilon)^2 + \taus^2}\right] \etavK,
    \label{eq:v_i,x,0}\\
    v_{i,y,0} &= -\left[1 - \frac{\taus^2}{(1 + \epsilon)^2 + \taus^2}\right]  \frac{\etavK}{1 + \epsilon},
    \label{eq:v_i,y,0}\\
    v_{i,z,0} &= 0.
    \label{eq:v__i,z,0}
\end{align}

\subsubsection{Pressureless Dust Fluid}
\label{sec:pressureless_dust_fluid}

For the density $\rhod$ and total velocity $\Vvec$ of dust modeled as a pressureless fluid, the inviscid continuity and momentum equations are \citep[cf.][eq.~2]{Pencil_dust}
\begin{align}
    \frac{\partial\rhod}{\partial t} + \nabla\cdot(\rhod\Vvec) = &\;0,
    \label{eq:dust_cont}\\
    \frac{\partial\rhod\Vvec}{\partial t} + \nabla\cdot(\rhod\Vvec\Vvec) = &\;\rhod\left[3\Omega^2x\xhat + 2\Vvec\times\bm{\Omega}\right.\nonumber\\
    &\left.-\,\frac{1}{\rhod}\left(\frac{\Vvec - \Uvec}{\tstop}\right)\right],
    \label{eq:dust_mom}
\end{align}
respectively.
The initial velocity components take the equilibrium solution by \citet{NakagawaSekiyaHayashi1986}, i.e., equations~\eqref{eq:v_i,x,0}--\eqref{eq:v__i,z,0} without the subscript $i$ for individual Lagrangian particles [cf. gas equations~\eqref{eq:u_x,0}--~\eqref{eq:u_z,0}].
The right-hand side of equation~\eqref{eq:dust_mom} parallels that of equation~\eqref{eq:gas_mom} without the radial gas pressure gradient (Footnote~\ref{foot:pressure_gradient}).

\subsubsection{Problem~BA} \label{sec:problem_ba}

This problem and its associated variations are based on the BA run from \cite{JohansenYoudin2007}.
As key parameters, $\Pi = 0.05$, $\epsilon = 0.2$, and $\taus = 1.0$, [equations~\eqref{eq:Pi}, \eqref{eq:epsilon}, and \eqref{eq:taus}, respectively].
Following their original setup, the domain size is $L_x \times L_z = 2H \times 2H$ (cf. their Tab.~1).
Although our fiducial grid resolution is $N_x \times N_z = 512 \times 512$, we also compare with simulations run at a higher resolution of $1024 \times 1024$.
Models are run until $\tlim= 100T$ where
\begin{equation}
    T \equiv 2\pi/\Omega,
    \label{eq:T}
\end{equation}
is the local orbital period.

\begin{deluxetable}{cccccl}
    \tablecaption{Parameters\label{tab:parameters}}
    \tablecolumns{6}
    \tablehead{
        \colhead{Problem}   & \colhead{$\taus$} & \colhead{$\epsilon$}  & \colhead{$L_{x,z}/H$} & \colhead{$N_x \times N_z$} & \colhead{$\np$}   \\
        (1)                 & (2)               & (3)                   & (4)                 & (5)                        & (6)}
    \startdata
        BA                  & 1.0               & 0.2                   & 2.0                 & $512 \times 512$           & $1, 9$            \\
                            &                   &                       &                     & $1024 \times 1024$         & $1$               \\
    \enddata
    \tablecomments{Columns are (1) problem name, (2) dimensionless stopping time$\anote$, (3) total dust-to-gas mass ratio$\bnote$, (4) domain size in gas scale heights, (5) grid resolution, and (6) average number of Lagrangian particles per grid cell/point.\\
        $\anote$ Defined by equation~\eqref{eq:taus}\\
        $\bnote$ Defined by equation~\eqref{eq:epsilon}}
\end{deluxetable}

These values and those that follow are summarized in Table~\ref{tab:parameters}.
For codes that model Lagrangian particles (Section~\ref{sec:lagrangian_dust_particles}), we initialize a uniformly random distribution of particle positions such that there is an average number of particles $n_\mathrm{p}$ per cell (or point; Section~\ref{sec:pencil}), with $n_\mathrm{p} = 1$ or 9 at our fiducial resolution but only $n_\mathrm{p} = 1$ at the higher resolution.
On the other hand, for codes that model a pressureless dust fluid (Section~\ref{sec:pressureless_dust_fluid}), the initial dust velocity components ($v_x, v_y$, and $v_z$) are perturbed with either uniform white noise on the order of $0.01\cs$ (in Idefix; Section~\ref{sec:idefix}) or Gaussian (i.e., normally distributed) noise with a standard deviation of $v_\mathrm{rms} = 0.01\cs$.

The decision to perturb the initial density or velocity field is influenced by the intrinsic differences between particle and fluid dust models (see Section~\ref{sec:discussion} for additional discussions).
The different noise used (e.g., Poisson, white, or Gaussian) may not guarantee the same magnitude of initial perturbations between fields, codes, or dust models, which may result in slight differences in their early evolution, e.g., $t \lessapprox 10T$ (see Figures~\ref{fig:BA-512_time_series_pre-10T}, \ref{fig:BA-512-np9_time_series_pre-10T}, and \ref{fig:BA-1024_time_series_pre-10T} and their corresponding descriptions in Sections~\ref{sec:fiducial_grid_resolution}, \ref{sec:higher_particle_resolution}, and \ref{sec:higher_grid_resolution}, respectively).
However, given the stochastic nature of this nonlinear problem (Section~\ref{sec:stochasticity}), the most robust comparisons are of the statistical properties of the saturated state, which all codes reach by $20T$.

\subsection{Codes and Numerical Methods} \label{sec:codes_numerical_methods}

This section introduces the various numerical techniques and hydrodynamic codes used to model the streaming instability for our comparative study.
Section~\ref{sec:comp_vels} briefly discusses the Galilean invariant algorithm that some of the codes here used, while the following sections introduce each code alphabetically and highlight their features, relevant integration methods, and available dust models.
Table~\ref{tab:codes} lists key aspects for quick reference and direct comparison between codes,
and Appendix~\ref{appx:public_repository} (Table~\ref{tab:repos}) provides the current URLs to their publicly available repositories.

\begin{deluxetable*}{lcllcllllll}
    \tablecaption{Codes\label{tab:codes}}
    \tablecolumns{10}
    \tablehead{
        \colhead{Code}  & \colhead{Section}     & \colhead{Dust}    & \colhead{Equations}                                                           & \multicolumn{2}{c}{Riemann Solver}      & \multicolumn{2}{c}{Reconstruction}            & \multicolumn{2}{c}{Time}          \\
                        &                       &                   &                                                    &\multicolumn{1}{c}{Gas}&\multicolumn{1}{c}{Dust}&\multicolumn{1}{c}{Gas}&\multicolumn{1}{c}{Dust}&\multicolumn{1}{c}{Integrator}&\multicolumn{1}{c}{Order}\\
        \multicolumn{1}{c}{(1)}&(2)             &\multicolumn{1}{c}{(3)}&\multicolumn{1}{c}{(4)}                                                                    &(5)&\multicolumn{1}{c}{(6)}&\multicolumn{1}{c}{(7)}&\multicolumn{1}{c}{(8)}&\multicolumn{1}{c}{(9)}&\multicolumn{1}{c}{(10)}}
    \startdata
        Athena          & \ref{sec:athena}      & Particles$\anote$ & \eqref{eq:gas_mom}, \eqref{eq:par_acc}                                        & HLLC$\cnote$          & \nodata           & PPM$\hnote$               & \nodata           & CTU$\mnote$               & 2nd   \\
        Athena++        & \ref{sec:athena++}    & Particles         & \eqref{eq:gas_mom_resid}, \eqref{eq:par_vel_resid}, \eqref{eq:par_acc_resid}  & HLLE$\dnote$          & \nodata           & PPM$\inote$               & \nodata           & VL2$\nnote$               & 2nd   \\
                        &                       & Fluid$\bnote$     & \eqref{eq:gas_mom}, \eqref{eq:dust_mom}                                       & Roe$\enote$           & Custom$\fnote$    & PPM                       & PLM               & VL2                       & 2nd   \\
        Idefix          & \ref{sec:idefix}      & Fluid             & \eqref{eq:gas_mom}, \eqref{eq:dust_mom}                                       & HLLC                  & HLL               & PLM$\jnote$-VLFL$\knote$  & PLM-VLFL          & TVD$\onote$-RK2$\pnote$   & 2nd   \\
        LA-COMPASS      & \ref{sec:la-compass}  & Fluid             & \eqref{eq:gas_mom}, \eqref{eq:dust_mom}                                       & HLLE                  & HLLE              & PLM                       & PLM               & TR-BDF2$\qnote$           & 2nd   \\
        PLUTO           & \ref{sec:pluto}       & Particles         & \eqref{eq:gas_mom}, \eqref{eq:par_acc}                                        & Roe                   & \nodata           & PPM                       & \nodata           & RK2                       & 2nd   \\
                        &                       & Fluid             & \eqref{eq:gas_mom}, \eqref{eq:dust_mom}                                       & Roe                   & Exact$\gnote$     & PLM-MC$\lnote$            & PLM-MC            & RK2                       & 2nd   \\
        \hline
        FARGO3D         & \ref{sec:fargo3d}     & Fluid             & \eqref{eq:gas_mom}, \eqref{eq:dust_mom}                                       & \nodata               & \nodata           & PLM-VLFL                  & PLM-VLFL          & Euler                     & 1st   \\
        \hline
        Pencil          & \ref{sec:pencil}      & Particles         & \eqref{eq:gas_mom_resid}, \eqref{eq:par_vel_resid}, \eqref{eq:par_acc_resid}  & \nodata               & \nodata           & \nodata                   & \nodata           & RK4                       & 4th
    \enddata
    \tablecomments{Columns give (1) the code name, (2) the section with its description, (3) its supported dust model(s); (4) the equations of motion solved, the Riemann solvers for the (5) gas and (6) dust fluid, the spatial reconstruction methods for the (7) gas and (8) dust fluid, and (9) the time integrator and (10) its order of accuracy used for this project.
    The upper group of codes all use the finite-volume method, while FARGO3D uses both finite-volume and finite-difference methods via operator splitting with conservative upwind, and Pencil uses only the finite-difference method.\\
        $\anote$ Section~\ref{sec:lagrangian_dust_particles}\\
        $\bnote$ Section~\ref{sec:pressureless_dust_fluid}\\
        $\cnote$ Harten--Lax--van Leer \citep{HLL} for contact waves \citep{HLLC2, HLLC1}\\
        $\dnote$ Harten--Lax--van Leer--Einfeldt \citep{HLLE}\\
        $\enote$ \citet{Roe}\\
        $\fnote$ \citet[][\S~2.4.1]{Athena++_fluid}\\
        $\gnote$ \cite{LeVeque2004}\\
        $\hnote$ Piecewise parabolic method \citep{PPM}\\
        $\inote$ Uses the \cite{McCorquodaleColella2011} flux limiter\\
        $\jnote$  Piecewise linear method\\
        $\knote$ van Leer flux limiter\\
        $\lnote$ Monotonized central difference limiter \citep{MC}\\
        $\mnote$ Corner transport upwind \citep{CTU}\\
        $\nnote$ van Leer \citep{vanLeer, VL2}\\
        $\onote$ Total variation diminishing\\
        $\pnote$ Runge--Kutta \citep{Runge1895, Kutta1901}\\
        $\qnote$ Trapezoidal rule and backward differential formula \citep{TR-BDF2}}
\end{deluxetable*}

\subsubsection{Differences in Computing Velocities} \label{sec:comp_vels}

Codes that solved equations~\eqref{eq:gas_mom}, \eqref{eq:par_acc}, or \eqref{eq:dust_mom} (Table~\ref{tab:codes}) used different dynamical variables to subtract the mean azimuthal flow in Keplerian disks, which in the 2D axisymmetric radial–vertical ($x$–$z$) problem considered here (Section~\ref{sec:problem_ba}) amounts to replacing the total velocities with their residuals [e.g., equations~\eqref{eq:tot_vel} and \eqref{eq:lin_kep_gas_vel}; Sections~\ref{sec:gas}–\ref{sec:pressureless_dust_fluid}].
Compared to solving equations~\eqref{eq:gas_mom_resid}, \eqref{eq:par_vel_resid}, or \eqref{eq:par_acc_resid} directly, we note that the algorithm can technically introduce computational overhead, although both are unnoticeable in practice.
Thus, we do not expect this to be a likely source of difference between the codes, i.e., with respect to their results (Section~\ref{sec:results}) or their performance (Section~\ref{sec:performance}).

\subsubsection{Athena} \label{sec:athena}

Athena \citep{Athena} is a second-order accurate Godunov flux-conservative (finite-volume) code designed to solve the equations of compressible magnetohydrodynamics (MHD).
\citet{StoneGardiner2010} first implemented the shearing-box approximation into the code.
It supports several approximate Riemann solvers, including the linear solver by \citet{Roe}, the Harten--Lax--van Leer \citep[HLL;][]{HLL} nonlinear solver for contact waves (HLLC\footnote{
`C' stands for \textit{Contact}.};
\citealt{HLLC2, HLLC1}) to compute time-averaged fluxes of all conserved hydrodynamic quantities at cell interfaces, and the HLLD\footnote{
`D' stands for \textit{Discontinuities} (specifically two additional MHD wave discontinuities: the Alfvén waves).}
solver for MHD \citep{HLLD1,HLLD2}.
The Athena models for this project use the HLLC solver, the dimensionally unsplit corner-transport-upwind \citep[CTU;][]{CTU} method for time integration, and the third-order piecewise parabolic method \citep[PPM;][]{PPM} for spatial reconstruction of the gas.

\citet{Athena_particles} first modeled the dust as Lagrangian particles in Athena.
The particle module adopts a predictor--corrector approach combined with the CTU integrator.
The predictor--corrector scheme integrates the momentum equations of the gas and particles coupled through the drag force \citep[][eqs.~7a--7c]{Athena_particles}.
A semi-implicit integrator is used in this project for the particle momentum equation as an intermediate step of the predictor--corrector scheme.
To calculate the drag force at the particle positions, the triangular-shaped-cloud (TSC) scheme interpolates gas velocities centered at each cell.
The scheme also interpolates particle momenta onto cell centers to calculate the dust back-reaction to the gas.
In the Athena models for this project, a constant inward force is applied to the particles as a proxy for the global radial gas pressure gradient.
Thus, the $2\Omega\Pi \cs\xhat$ term moves from the gas momentum to the particle acceleration and becomes negative \citep[cf.][eq.~4]{Athena_particles} which is mathematically equivalent to equations~\eqref{eq:gas_mom_resid} and \eqref{eq:par_acc_resid}.

\subsubsection{Athena++} \label{sec:athena++}

Athena++ \citep{Athena++} is a complete rewrite of Athena (Section~\ref{sec:athena}) in C++.
It features more flexible coordinate options, including mesh refinement, and improved modularity, performance, and scalability with hybrid support for the OpenMP\footnote{
The OpenMP application programming interface (API) supports shared-memory multiprocessing through multi-threading but standalone is limited to as many cores available in a single multicore processor or in multi-socket node in a high-performance-computing cluster (cf. MPI; Footnote~\ref{foot:mpi}).\label{foot:openmp}}
and Message Passing Interface (MPI; also supported by the other codes)\footnote{
The Message Passing Interface (MPI) standard can facilitate distributed-memory parallelization between CPU cores and across multiple nodes in a high-performance-computing cluster (cf. OpenMP; Footnote~\ref{foot:openmp}).\label{foot:mpi}}
parallelization standards.
The Athena++ models for this project use PPM reconstruction
for the gas and the second-order van Leer predictor–corrector time integrator \citep{vanLeer,VL2}.

C.-C. Yang et al. (in preparation) first modeled the dust as Lagrangian particles in Athena++.
To integrate individual particle positions and velocities in unison with hydrodynamic time steps, the standard particle--mesh method \citep{HockneyEastwood1981}, using the TSC scheme, interpolates the gas properties onto the particles and assigns the particle properties to the mesh with high spatial accuracy.
These models use the Harten--Lax--van Leer--Einfeldt \citep[HLLE;][]{HLLE} Riemann solver for the gas.

\cite{Athena++_fluid} first modeled the dust as a pressureless fluid in Athena++.
The fluid approximation is generally valid when the dust stopping time is short, i.e., $\taus \lesssim 1$, which minimizes dust-crossing behavior \citep{GaraudBarriere-FouchetLin2004}.
Despite its limitations, the multi-fluid approach offers several advantages, e.g., the dust naturally shares the same spatial resolution as the gas when mesh refinement is enabled, and it provides automatic load balancing compared to the particle-mesh method.
Moreover, it allows for straightforward implementation of sub-grid-level turbulent effects, fully implicit schemes for dust-gas drag, and dust coagulation.
These dust fluid models use the Roe Riemann solver for the gas, a custom solver for the dust \citep[][\S~2.4.1]{Athena++_fluid}, and the piecewise linear method (PLM) to reconstruct the dust.
Using this multi-fluid module of Athena++, \citet{HuangBai2025ApJL} presented the first study of SI–driven dust clumping in a self-consistent turbulent environment within a 3D global framework.
Their results emphasize the necessity of a realistic treatment of hydrodynamic instabilities for modeling dust concentration and planetesimal formation.

\subsubsection{FARGO3D} \label{sec:fargo3d}

FARGO3D \citep{FARGO3D} is a 3D Eulerian MHD code, belonging to the ZEUS-type family and utilizing finite-difference upwind, dimensionally split methods \citep{StoneNorman1992}.
The public version also incorporates nonideal MHD \citep{KrappGresselBenitez-Llambay2018}, multi-fluid dust dynamics \citep{FARGO3D_fluid}, and a generalization of the advection algorithm \citep{Masset2000} for arbitrary meshes \citep{Benitez-LlambayKrappRamos2023}.
The code strictly conserves mass and momentum, satisfying Rankine--Hugoniot (shock--jump) conditions in barotropic systems.
It is designed for central processing units (CPUs) and graphics processing units (GPUs) in high-performance-computing clusters, with shared or distributed memory (Footnotes~\ref{foot:openmp} and \ref{foot:mpi}),
and written in C with Python scripts and a C-to-CUDA parser.\footnote{
CUDA (Compute Unified Device Architecture) is a proprietary API and parallel computing platform for NVIDIA GPUs.\label{foot:cuda}}
FARGO3D solves equations in Cartesian, cylindrical, or spherical coordinates on a static and staggered mesh, employing first-order time and second- or third-order upwind method for the flux integration. 
It uses operator splitting to divide a full time step into three sub-steps:
(1) a source step: finite-difference momentum and internal energy update with pressure, gravity, and viscosity;
(2) a collision step: momentum update through drag-force;
and (3) a transport step: conservative density, energy, and momentum advection using finite volumes and dimensional splitting with van Leer slope \citep{vanLeer} reconstruction for cell-centered flux interpolation to cell faces.
Users can include any number of point-like masses orbiting a central mass that can interact with the surrounding gas.
Planetary systems are evolved using the fifth-order Cash--Karp \citep{CashKarp1990} Runge–Kutta \citep{Runge1895,Kutta1901} method with a fixed time step.

\cite{FARGO3D_fluid} presented a novel numerical method designed to accurately capture momentum transfer in systems with multiple interacting fluid species.
The method exhibits asymptotic and unconditional stability and conserves total momentum to within machine precision.
The work also included a comprehensive multispecies test suite, applicable to an arbitrary number of species, based on analytical or exact solutions for problems encompassing perfect damping, damped sound waves, shocks, local and global gas--dust radial drift in a disk, and the linear streaming instability.
The method updates the momentum due to the drag force with a first-order implicit time integration.
In the case of gas and multi-fluid dust, the time complexity of the implicit method scales linearly with the number of fluids \citep{KrappBenitez-Llambay2020}, dramatically reducing the computational cost while preserving numerical stability.
\cite{KrappGarrido-DeutelmoserBenitez-Llambay2024} has extended the method to second-order and adapted it to implicit--explicit (IMEX) type schemes.

\subsubsection{Idefix} \label{sec:idefix}

Idefix \citep{Idefix} is a performance-portable finite-volume simulation code designed for computational fluid dynamics, based on a high-order Godunov scheme.
Although it adopts a structural design similar to the PLUTO code (Section~\ref{sec:pluto}), Idefix is implemented in C++ and employs the Kokkos programming model \citep{Kokkos} to enable efficient execution across multiple architectures, including both CPUs and GPUs.
The code supports a variety of grid geometries with nonuniform spacing and offers extensive flexibility through a wide selection of Riemann solvers, spatial reconstruction methods, and additional physics (nonideal MHD, self-gravity, orbital advection, etc.).

In the current implementation, dust is treated as a pressureless fluid, following the approach described by \cite{Idefix_dust}, and evolves using an HLL-type Riemann solver.
The gas component is modeled using the HLLC Riemann solver, with PLM spatial reconstruction based on the van Leer slope limiter.
Time integration is performed using a second-order total variation diminishing (TVD) Runge--Kutta (RK2) scheme.

\subsubsection{LA-COMPASS} \label{sec:la-compass}

The Los Alamos COMPutional Astrophysics Simulation Suite (LA-COMPASS) is a collection of several modern, high-resolution, Godunov-type (finite-volume) hydrodynamic and MHD codes that have been developed at Los Alamos National Laboratory using Laboratory Directed Research and Development funding.
As part of LA-COMPASS, the simulation code for protoplanetary disks was specifically developed for disk--planet interaction problems \citep{LA-COMPASS}.
It is fully parallelized and optimized through a hybrid of MPI and OpenMP standards for large-scale parallel computers with multiple cores.
The protoplanetary disk is treated as a 3D gas whose motion is described by the Navier--Stokes equations.
It also contains a modified Galilean-invariant FARGO scheme \citep[Section~\ref{sec:comp_vels};][]{Masset2000} and a fast self-gravity solver, although neither are used for this project.

\cite{LA-COMPASS_fluid} first modeled the dust as a pressureless fluid in LA-COMPASS.
Its evolution is governed by conservation laws, and the coupling between the gas and dust is mediated through the drag force, which is very stiff for short stopping times.
For both gas and dust, the LA-COMPASS models for this project use the HLLE Riemann solver, a second-order time integrator based on the trapezoidal rule and the backward differential formula \citep{TR-BDF2}, and PLM for reconstruction.

\subsubsection{Pencil} \label{sec:pencil}

A flagship code for streaming instability studies, Pencil \citep{BrandenburgDobler2002, Pencil} is a nonconservative high-order finite-difference code primarily designed for weakly compressible hydromagnetic turbulence.
For spatial derivatives, the user has choices of 2nd, 4th, 6th, 8th, and 10th-order accuracy, as well as centered, single-sided, or upwind.
For time integration, third (RK3) and fourth (RK4) order Runge--Kutta time-stepping schemes are implemented.
For this work, we use sixth-order centered spatial derivatives and RK4, except for a single run with RK3 (Table~\ref{tab:performance}, Note~f).
Because the high-order scheme has little dissipation overall, hyperviscosity is often employed to prevent unphysical accumulation of power at the grid scale, while keeping the integral scales inviscid \citep{HaugenBrandenburg2004, LyraJohansenKlahr2008, LyraMcNallyHeinemann2017};
shocks are treated with a von Neumann-type shock viscosity.

Although it was first used for planet formation in the fluid approximation \citep{JohansenAndersenBrandenburg2004}, Pencil has an advanced Lagrangian particle module, first implemented by \cite{Pencil_dust}.
A TSC scheme evaluates and assigns the drag force by interpolating the gas velocity using either a quadratic spline or a quadratic polynomial with respect to neighboring grid points (not cells).
Load balancing is achieved through block domain decomposition \citep[][\S~3.3]{JohansenKlahrHenning2011}.

\subsubsection{PLUTO} \label{sec:pluto}

PLUTO \citep{PLUTO, PLUTOII} is a high-order, Godunov-type shock-capturing, finite-volume code designed to model astrophysical plasmas. 
The code supports large-scale parallel computations through the MPI and Chombo \citep{Chombo} libraries with different grids (i.e., static or with adaptive mesh refinement).
PLUTO includes comprehensive physics and solvers with ideal or non-ideal (relativistic) MHD, multiple particle modules [e.g., MHD-PIC (particle-in-cell) and Lagrangian particles, \citealt{MignoneBodoVaidya2018}], radiation hydrodynamics with the M1 moment method \citep{MelonFuksmanMignone2019, MelonFuksmanKlahrFlock2021}, and the FARGO scheme \citep{PLUTO_FARGO}.
There is active development of a fourth-order finite-volume scheme \citep{BertaMignoneBugli2024}, and a high-order GPU-accelerated version (gPLUTO) containing most of the key features of the legacy CPU version.
All PLUTO models for this project use the Roe Riemann solver and the RK2 time integrator for the gas.
    
The current version supports solving for the motion of dust using Lagrangian particles \citep{PLUTO_particles}.
It includes time integration using the exponential midpoint method by \cite{HochbruckOstermann2010} and the semi-implicit scheme used in Athena (see \ref{sec:athena}), with the exponential midpoint method providing several advantages over the latter—time reversibility, bounded energy errors, and better stability.
Particle weighting is done using the TSC scheme.
For the runs here, the particle models use fourth-order PPM reconstruction for the gas.

PLUTO is also able to model dust as a pressureless fluid using both explicit and implicit schemes \citep{PLUTO_fluid}, with a multi-fluid version under development (Sudarshan et al., in preparation). 
The dust fluid models here use an exact Riemann solver \citep{LeVeque2004} for the dust, a second-order PLM to reconstruct the gas, and an explicit integrator.

\section{Results} \label{sec:results}

We compare results from the codes and methods described in Section~\ref{sec:codes_numerical_methods} for Problem~BA and its variations specified in Section~\ref{sec:problem_ba}.
Sections~\ref{sec:fiducial_grid_resolution} and \ref{sec:higher_grid_resolution} report runs at our fiducial and higher grid resolutions, respectively, and Section~\ref{sec:performance} compares the computational performance of these and additional runs.

\subsection{Fiducial Grid Resolution} \label{sec:fiducial_grid_resolution}

\begin{figure*}
    \centering
    \includegraphics[scale=1.044]{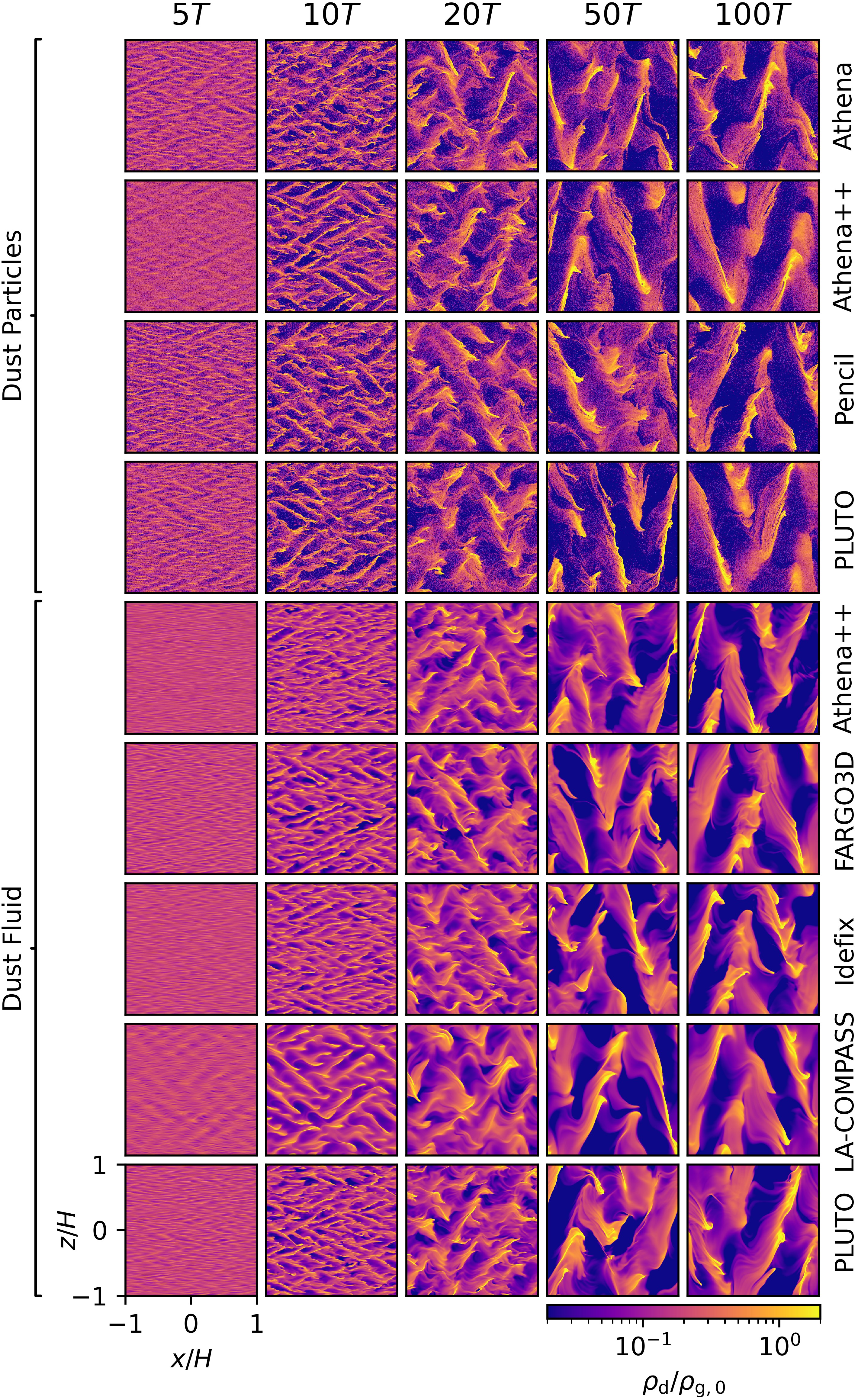}
    \caption{Dust density snapshots for Problem~BA at grid resolution $512^2$.
    Each row shows those from a single code (Section~\ref{sec:codes_numerical_methods}), and each column shows the simulation time in local orbital periods $T$.
    In alphabetical order from top to bottom, upper and lower groups implement an average of $\np = 1$ particle per grid cell/point and a pressureless dust fluid, (Sections~\ref{sec:lagrangian_dust_particles} and \ref{sec:pressureless_dust_fluid}) respectively.
    Radial $x$ and vertical $z$ coordinates are in units of the vertical gas scale height $H$.
    The color scale at the bottom right shows the dust density $\rhod$ relative to the initially uniform gas density $\rhogn$ and applies to all snapshots.}
    \label{fig:BA-512_snapshots}
\end{figure*}

For our fiducial grid resolution of $512^2$ (Section~\ref{sec:problem_ba}), Figure~\ref{fig:BA-512_snapshots} shows a series of dust density $\rhod$ snapshots.
In general, each system follows a similar evolution: perturbations---seen as a regular pattern of short, alternatingly slanted filaments---grow exponentially ($5 < t/T < 10$) before reaching a saturated state as the filaments begin to roll and break ($10 < t/T < 20$);
as the remaining filaments continue to merge with each other, the density of the filaments gradually increases until a steady state is achieved ($20 < t/T < 100$).
To further visualize and demonstrate this evolution, a public collection of videos, with high frame rates at various grid resolutions, is maintained on Figshare\footnote{
Dust density evolution videos:
\url{http://doi.org/10.6084/m9.figshare.c.6718221}.\label{foot:videos}}\citep{BaronettYangZhu2024_videos}.
Although the snapshots taken at the same orbital time from different codes look qualitatively similar overall, closer inspection of the
upper group of four reveals expected Poisson noise associated with a Lagrangian implementation of particles (e.g., the first PLUTO and second Athena++ rows side by side).

\begin{figure}
    \includegraphics[width=\columnwidth]{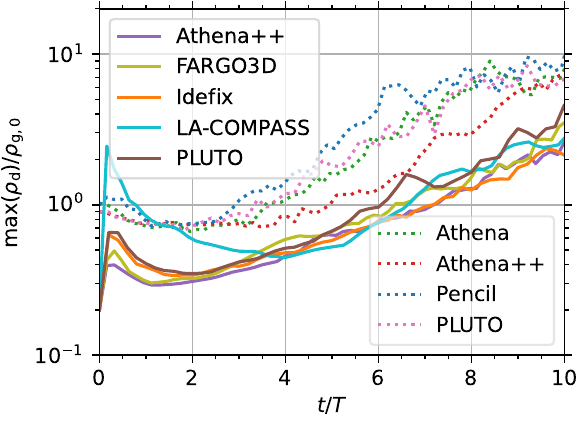}
    \caption{Maximum dust density $\max(\rhod)$ as a function of time $t < 10T$ for Problem~BA at grid resolution $512^2$.
    The units for density and time are the initially uniform gas density $\rhogn$ and the local orbital period $T$, respectively.
    Using a similar cadence for each time series, dotted lines show particle codes (with average of $\np = 1$ particle per grid cell/point), solid lines show codes with pressureless dust fluids, and colors differentiate the codes (Sections~\ref{sec:lagrangian_dust_particles}, \ref{sec:pressureless_dust_fluid}, and \ref{sec:codes_numerical_methods}, respectively).}
    \label{fig:BA-512_time_series_pre-10T}
\end{figure}

Figure~\ref{fig:BA-512_time_series_pre-10T} helps to distinguish the early evolution of each code.
After the first time step in all models, $\max(\rhod)$ decays by less than an order of magnitude to a local minimum before turning over to exponential growth by $t \lessapprox 5T$.
For codes with particles (dotted lines), the local minima reach $\max(\rhod) \approx 0.7\rhogn$ approximately between $2T$ (for Athena, Pencil, and PLUTO) and $3T$ (for Athena++).
Except for LA-COMPASS, codes with dust fluids (solid lines) reach local minima $\max(\rhod) \approx 0.3\rhogn$ as early as $1T$ (for Athena++ and FARGO3D) or $2T$ (for Idefix and PLUTO).
Between dust models, the factor of two difference in $\max[\rhod(t < 10T)]$ and the $\Delta t \sim 1T$ difference in the start of exponential growth may be due to the difference in initial perturbations, i.e., density noise for particles versus velocity noise for fluids (Section~\ref{sec:problem_ba}).
We note that LA-COMPASS shows the largest initial decay, from $\max(\rhod) \approx 2\rhogn$ to $\max(\rhod) \approx 0.4\rhogn$, over the longest window, $0 \lessapprox t/T \lessapprox 4$.

\begin{figure}
    \includegraphics[width=\columnwidth]{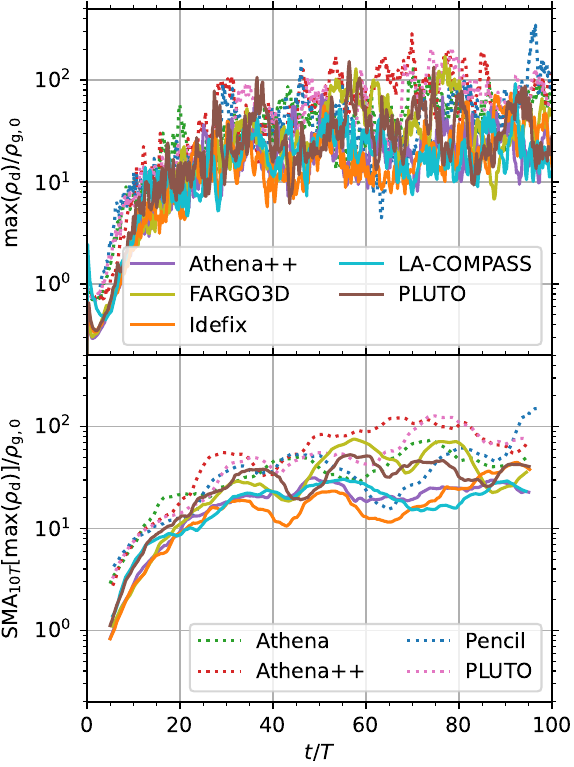}
    \caption{Similar to Figure~\ref{fig:BA-512_time_series_pre-10T} except until $\tlim = 100T$.
    The lower panel shows the simple moving average (SMA) with a sampling window of $10T$.}
    \label{fig:BA-512_time_series}
\end{figure}

Figure~\ref{fig:BA-512_time_series} shows the entire time series of $\max(\rhod)$ to $\tlim = 100T$.
In the upper panel, the similarly high cadence does not show any obvious differences in time variability between the codes or dust models.
Unlike the early differences seen up to $10T$ in Figure~\ref{fig:BA-512_time_series_pre-10T}, the knees of $\max(\rhod)$ between $10 < t/T < 15$, which mark the end of exponential growth, appear to be less dependent on the dust model.
Moreover, all models have reached a saturated state by $20T$.

In the lower panel of Figure~\ref{fig:BA-512_time_series}, the simple moving averages ($\SMA$), with a sampling window of $10T$, begin to highlight the overall difference in $\max(\rhod)$ between the dust models.
At a resolution of $512^2$, codes that model dust with $\np = 1$ particles tend to maintain a higher $\max(\rhod)$ than those that model it as a fluid.
Moreover, $\SMA_{10T}[\max(\rhod)]$ reaches as high as $10^2\rhogn$ for particles (e.g., Athena++, Pencil, and PLUTO) but stays as low as $\sim$$10\rhogn$ for fluids (e.g., Idefix) during the saturated state (e.g., $t > 30T$).

\begin{figure}
    \includegraphics[width=\columnwidth]{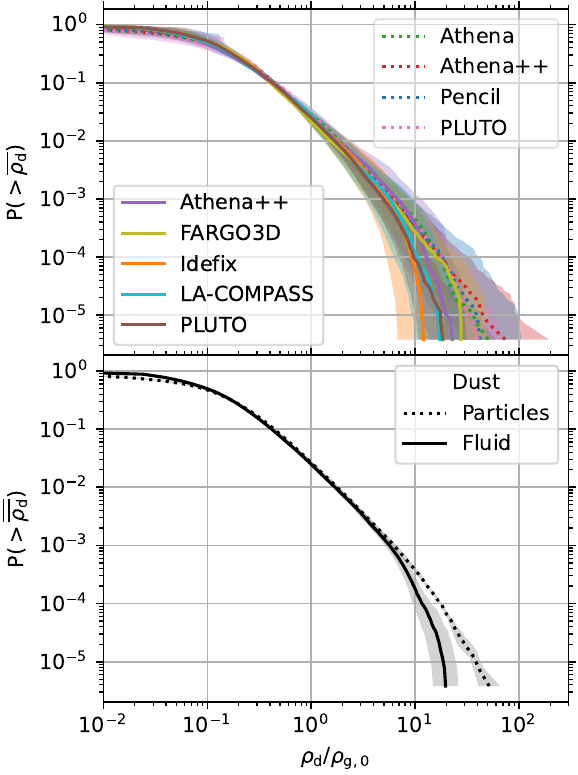}
    \caption{Cumulative distribution functions of the dust density $\rhod$ for Problem~BA at grid resolution $512^2$.
    Dotted and solid lines show codes that implement an average of $\np = 1$ particle per grid cell/point and a pressureless dust fluid, (Sections~\ref{sec:lagrangian_dust_particles} and \ref{sec:pressureless_dust_fluid}) respectively, with all $\rhod$ relative to the initially uniform gas density $\rhogn$.
    In the upper panel, curves show time averages over the saturated state $\avgrhod$ (cf. Figure~\ref{fig:BA-512_time_series}), colors different codes (Section~\ref{sec:codes_numerical_methods}), and shaded areas the $1\sigma$ time variability.
    In the lower panel, curves show the mean $\meanavgrhod$ of all particle or fluid codes in the upper panel, and shaded areas the standard deviation of their respective $\avgrhod$.}
    \label{fig:BA-512_CDF}
\end{figure}

The cumulative distribution functions of dust density $\Prob(>\avgrhod)$ in the upper panel of Figure~\ref{fig:BA-512_CDF} better distinguish each code throughout the saturated state, where dust density fields $\avgrhod$ are averaged over time $t \geq 20T$.
Taking into account their $1\sigma$ time variability throughout the saturated state (shaded areas), the upper 99\% of each distribution ($\Prob > 10^{-2}$) is in good agreement with each other for $\rhod \lessapprox 2\rhogn$, which verifies the qualitative similarities between the snapshots at 20, 50, and $100T$ in Figure~\ref{fig:BA-512_snapshots} with its maximum color scale of $2\rhogn$.
However, $\Prob \lesssim 10^{-3}$ diverges by up to an order of magnitude between codes at $\max(\avgrhod)$ [cf. Idefix (solid orange line) and Athena++ with particles (dotted red line)].
To compare directly with dust fluid codes, we use the Lagrangian particle densities interpolated at grid centers (Section~\ref{sec:lagrangian_dust_particles}).
However, we note that the cumulative distributions of Lagrangian particles throughout this study (e.g., Sections~\ref{sec:higher_particle_resolution} and \ref{sec:higher_grid_resolution}) are consistently steeper than those found by \citet[][Fig.~6, bottom panels]{Athena_particles}, who measured densities at the particle locations (not grid centers).

The lower panel of Figure~\ref{fig:BA-512_CDF} distinguishes the density distributions throughout the saturated state by dust model, where $\meanavgrhod$ is the mean of all particle (dotted lines) or fluid (solid lines) codes shown in the upper panel.
The divergence of $\Prob(>\meanavgrhod) \lessapprox 3\times10^{-4}$ exceeds one standard deviation of the $\avgrhod$ value for each dust model (fluid vs. particles; gray-shaded areas), with up to 2.6$\times$ greater $\max(\meanavgrhod)$ for particles than dust as a fluid.
Although the shaded areas in this lower panel do not capture the wide time variabilities of the codes throughout the saturated state, their narrowness (i.e., smaller standard deviations) are consistent with the visual clustering of codes by dust model seen at higher end densities in the upper panel.
We further discuss the differences between the dust models in Section~\ref{sec:dust_models}.

\subsubsection{Higher Particle Resolution} \label{sec:higher_particle_resolution}

\begin{figure*}
    \centering
    \includegraphics[scale=1.044]{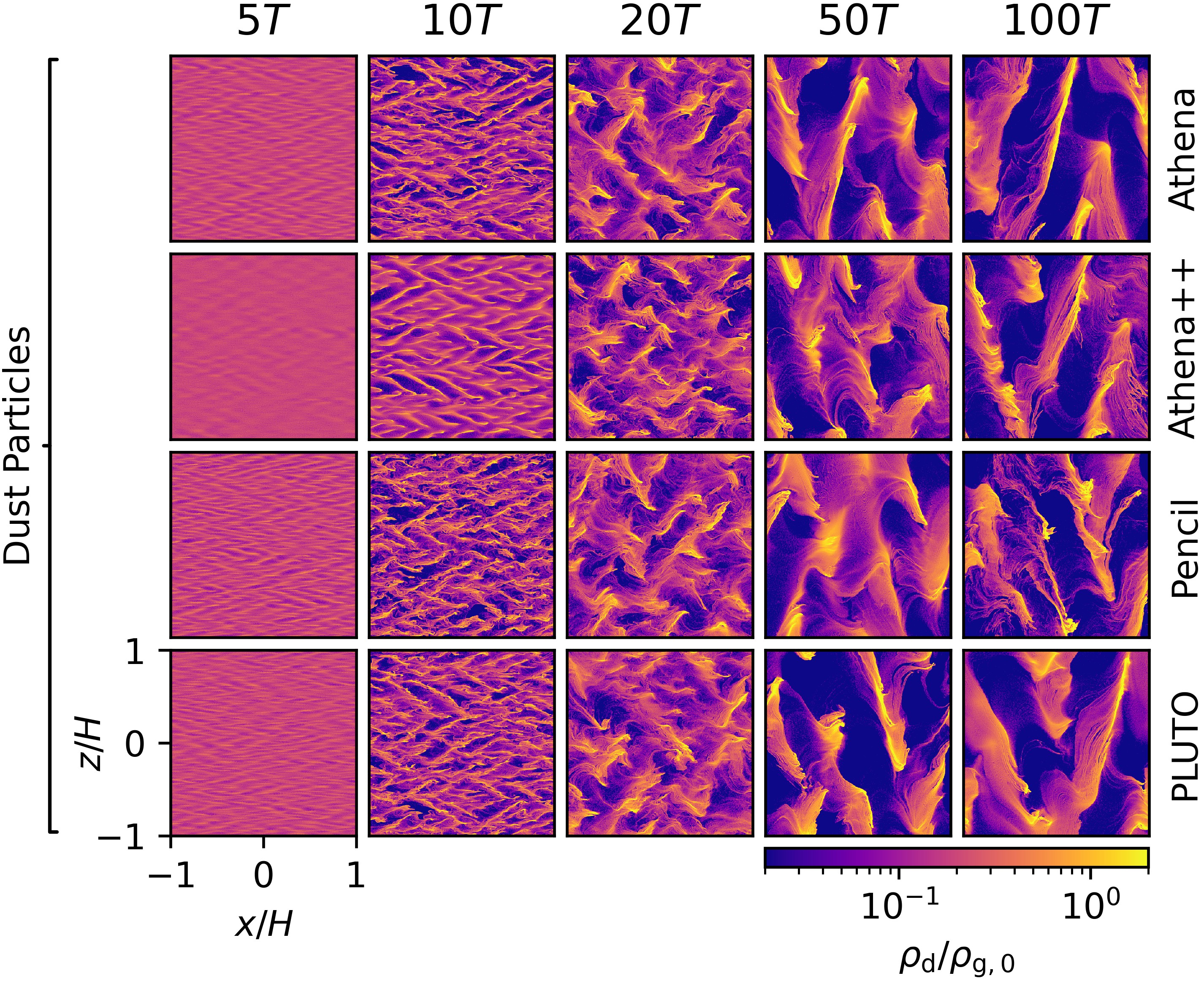}
    \caption{Similar to Figure~\ref{fig:BA-512_snapshots} except for particle codes with $\np = 9$ only.}
    \label{fig:BA-512-np9_snapshots}
\end{figure*}

Somewhat analogous to aspects of smoothed-particle hydrodynamics, increasing the number of Lagrangian particles (Section~\ref{sec:lagrangian_dust_particles}) effectively increases the mass resolution of the dust. 
Similar to Figure~\ref{fig:BA-512_snapshots}, Figure~\ref{fig:BA-512-np9_snapshots} shows snapshots of $\rhod$ from the various codes running at our fiducial grid resolution of $512^2$ (Section~\ref{sec:fiducial_grid_resolution}) except with an average of $\np = 9$ particles per grid cell/point instead of one (Section~\ref{sec:problem_ba}).
Close comparison of Figure~\ref{fig:BA-512-np9_snapshots} with the top four rows of Figure~\ref{fig:BA-512_snapshots} reveals the noticeable reduction in Poisson noise achieved by resolving the same total dust mass across nine times more particles, as expected.

\begin{figure}
    \includegraphics[width=\columnwidth]{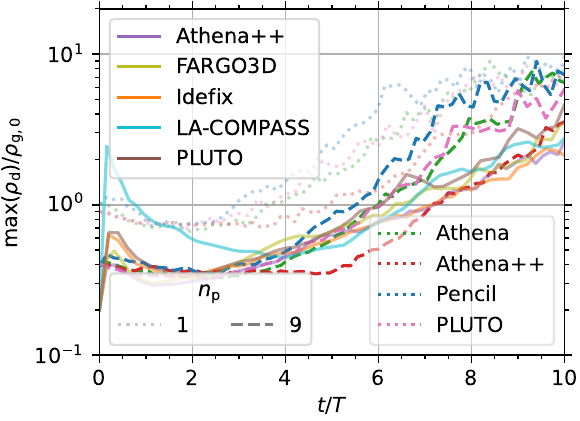}
    \caption{Figure~\ref{fig:BA-512_time_series_pre-10T} replotted with faded lines and with dashed lines showing $\np = 9$.}
    \label{fig:BA-512-np9_time_series_pre-10T}
\end{figure}

Figure~\ref{fig:BA-512-np9_time_series_pre-10T} shows the early evolution of these $\np = 9$ runs (dashed lines) compared to those previously shown in Figure~\ref{fig:BA-512_time_series_pre-10T} (faded lines) from Section~\ref{sec:fiducial_grid_resolution}.
Overall, the $\max[\rhod(t < 10T)]$ at higher particle resolutions almost match those for dust fluids (excluding LA-COMPASS).
The reduction in the rms of the initial noise perturbations (Section~\ref{sec:problem_ba}) for $\np = 9$ (by a factor of three compared to $\np = 1$; cf. Figures~\ref{fig:BA-512_snapshots} and \ref{fig:BA-512-np9_snapshots}), results in a significant reduction in the magnitudes of $\max[\rhod(t < 4T)]$ (by more than a factor of two) to the level of those for fluid codes.
For $t > 4T$, the exponential growth of models with $\np = 9$ is slightly steeper than that for dust fluids (but not compared to $\np = 1$ particles for $t > 2T$), although Athena++ with $\np = 9$ is again delayed compared to the other particle codes by about $2T$ (cf. Figure~\ref{fig:BA-1024_time_series_pre-10T}; see Section~\ref{sec:dust_models} for further discussion).

\begin{figure}
    \includegraphics[width=\columnwidth]{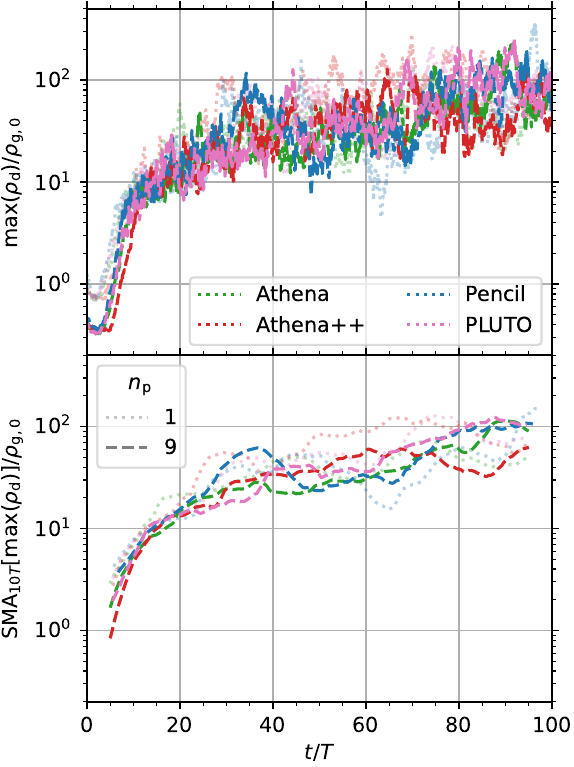}
    \caption{Similar to Figure~\ref{fig:BA-512_time_series} except only for particle codes, with dashed lines showing $\np = 9$ and faded dotted lines $\np = 1$.}
    \label{fig:BA-512-np9_time_series}
\end{figure}

\begin{figure}
    \includegraphics[width=\columnwidth]{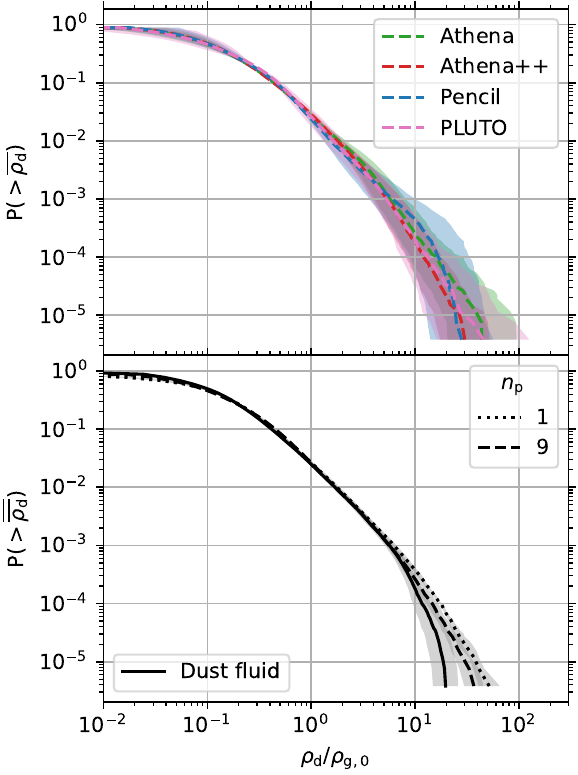}
    \caption{Similar to Figure~\ref{fig:BA-512_CDF}, except the upper panel only shows particle codes with $\np = 9$, and the lower panel shows their mean $\meanavgrhod$ with a dashed black curve.}
    \label{fig:BA-512-np9_CDF}
\end{figure}

Similar to Figure~\ref{fig:BA-512_time_series}, Figure~\ref{fig:BA-512-np9_time_series} shows the entire time series except only for particle codes at $512^2$ with $\np = 1$ and 9.
For $t > 15T$, there are few obvious differences between codes, but we note the appearance of a slightly more limited range in the time variability of $\max(\rhod)$ during the saturated state for $\np = 9$ (cf. the upper panels of Figures~\ref{fig:BA-512_CDF} and \ref{fig:BA-512-np9_CDF}).
Nonetheless, we find similar maximum densities with respect to particle resolution.

Similar to Figure~\ref{fig:BA-512_CDF}, Figure~\ref{fig:BA-512-np9_CDF} distinguishes the average dust density distributions throughout the saturated state.
In the lower panel, the means $\meanavgrhod$ of the time-averaged $\avgrhod$ of codes with either $\np = 1$ or 9 (upper panel) agree well for $\rhod < 10\rhogn$.
However, for $\rhod > 10\rhogn$, the two continue to diverge to just beyond one standard deviation of each other.
At $\max(\meanavgrhod)$, $\np = 9$ is slightly lower than $\np = 1$ and closer to that for dust fluids at $512^2$ grid resolution.
This decrease in the highest densities is consistent with a similar study of $\np$ by \citet[][Fig.~6, lower right panel]{Athena_particles} for Problem~BA at a resolution of $256^2$.

\subsection{Higher Grid Resolution} \label{sec:higher_grid_resolution}

\begin{figure*}
    \centering
    \includegraphics[scale=1.044]{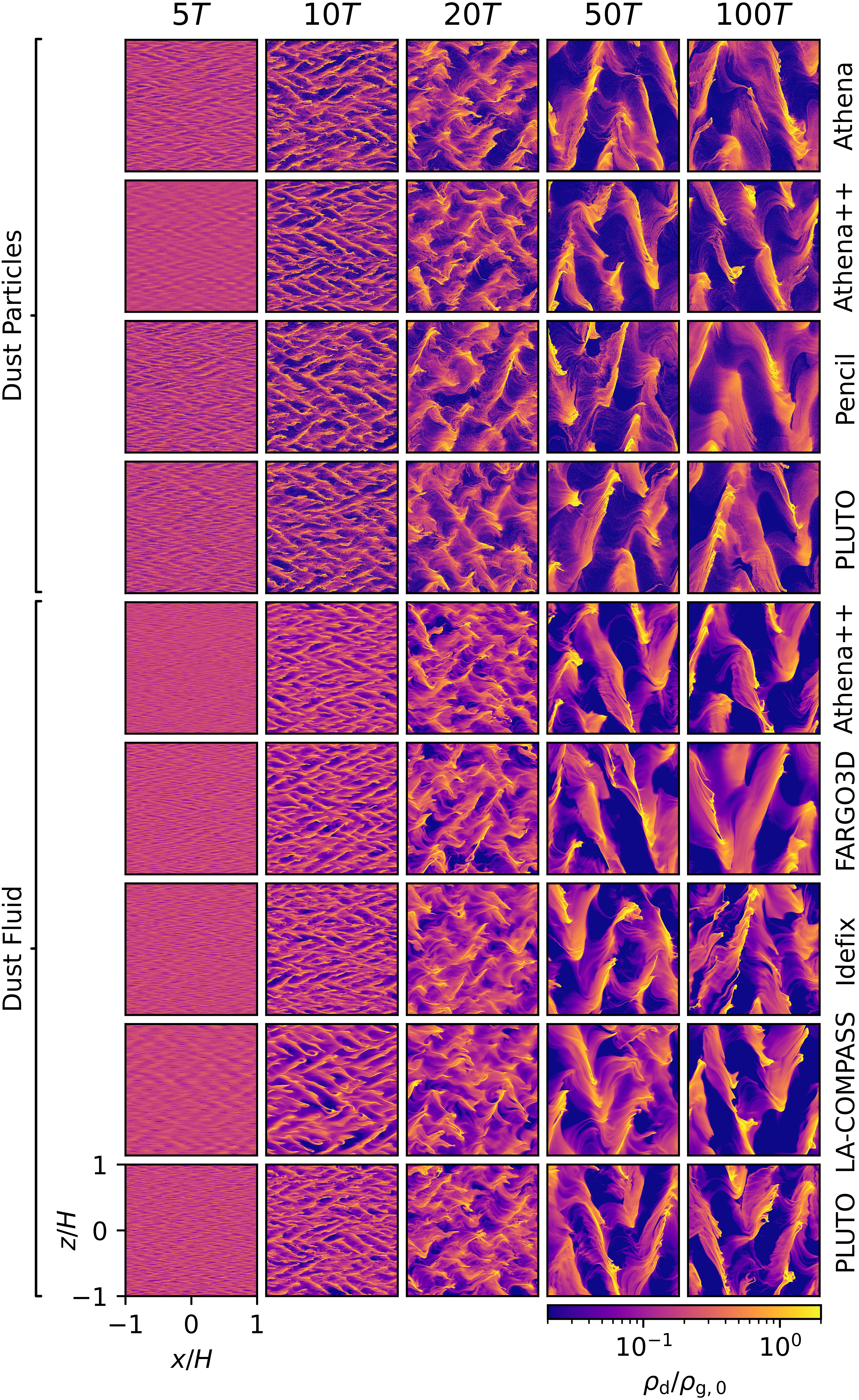}
    \caption{Similar to Figure~\ref{fig:BA-512_snapshots} except at grid resolution $1024^2$.}
    \label{fig:BA-1024_snapshots}
\end{figure*}

To investigate how spatial resolution affects each code and dust model, we compare runs at our fiducial grid resolution of $512^2$ with those at $1024^2$.
Figure~\ref{fig:BA-1024_snapshots} is similar to Figure~\ref{fig:BA-512_snapshots} but shows snapshots of the dust density $\rhod$ only from these higher-resolution runs.
Keeping an average of $\np = 1$ particles per grid cell/point (Section~\ref{sec:problem_ba}), the Lagrangian dust codes here feature four times as many particles as those in Section~\ref{sec:fiducial_grid_resolution}.
As in the case of $\np = 9$ at $512^2$ (Section~\ref{sec:higher_particle_resolution}; cf. Figure~\ref{fig:BA-512-np9_snapshots}), these additional particles (along with the finer grid) help reduce the Poisson noise seen in the top four rows of Figure~\ref{fig:BA-512_snapshots}.
Aside from this, the snapshots at a given orbital time $t/T$ overall look similar between the figures.

\begin{figure}
    \includegraphics[width=\columnwidth]{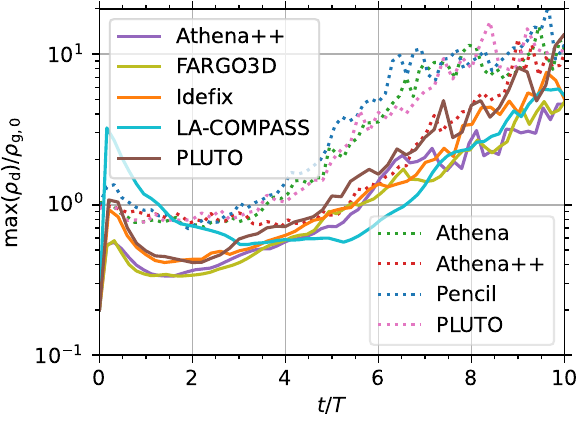}
    \caption{Similar to Figure~\ref{fig:BA-512_time_series_pre-10T} except at grid resolution $1024^2$.}
    \label{fig:BA-1024_time_series_pre-10T}
\end{figure}

Figure~\ref{fig:BA-1024_time_series_pre-10T} helps distinguish $\max[\rhod(t \leq 10T)]$ for these higher resolution models.
In general, their early evolution resembles that at $512^2$ (cf. Figure~\ref{fig:BA-512_time_series_pre-10T}) except for a couple of differences (cf. Section~\ref{sec:fiducial_grid_resolution}).
First, the exponential growth in $\max[\rhod(t > 4T)]$ is slightly steeper (e.g., five out of nine models surpass $10\rhogn$ before $10T$), as the higher spatial resolution can resolve more faster-growing linear modes \citep[][Fig.~1, upper-left panel]{Pencil_dust}.
Second, the gap between dust models almost closes by $10T$ and is narrower overall, as the different approaches to initializing perturbations become less important for the nonlinear evolution at a higher resolution (Section~\ref{sec:problem_ba}).

\begin{figure}
    \includegraphics[width=\columnwidth]{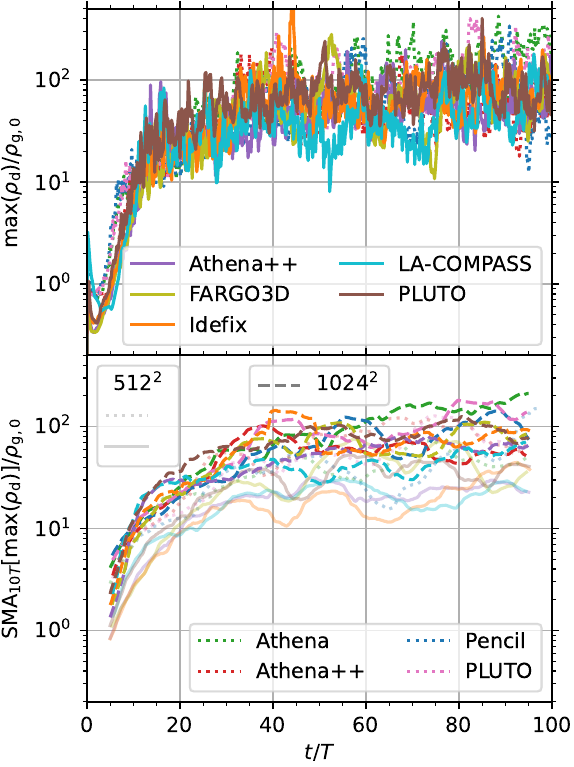}
    \caption{Similar to Figure~\ref{fig:BA-512_time_series}, except the upper panel only shows grid resolutions at $1024^2$, and the lower panel shows them as dashed lines along with faded line styles for $512^2$.}
    \label{fig:BA-1024_time_series}
\end{figure}

Similar to Figure~\ref{fig:BA-512_time_series}, Figure~\ref{fig:BA-1024_time_series} compares $\max[\rhod(t \leq 100T)]$.
In the upper panel, the range in the time variability of $\max[\rhod(t)]$ is reduced to about an order of magnitude compared to the wider range seen between codes at $512^2$ in Figure~\ref{fig:BA-512_time_series}.
For Idefix, we note a very brief order-of-magnitude jump to almost $2000\rhogn$ in $\max[\rhod(t \approx 44T)]$.
In the lower panel, the 10-orbit simple moving average generally shows higher $\max(\rhod)$ throughout the saturated state ($t > 20T$) for $1024^2$ (dashed lines) than for $512^2$ (faded lines), consistent with the resolution study by \citet[][Fig.~A2, upper right panel]{BaronettYangZhu2024}.

\begin{figure}
    \includegraphics[width=\columnwidth]{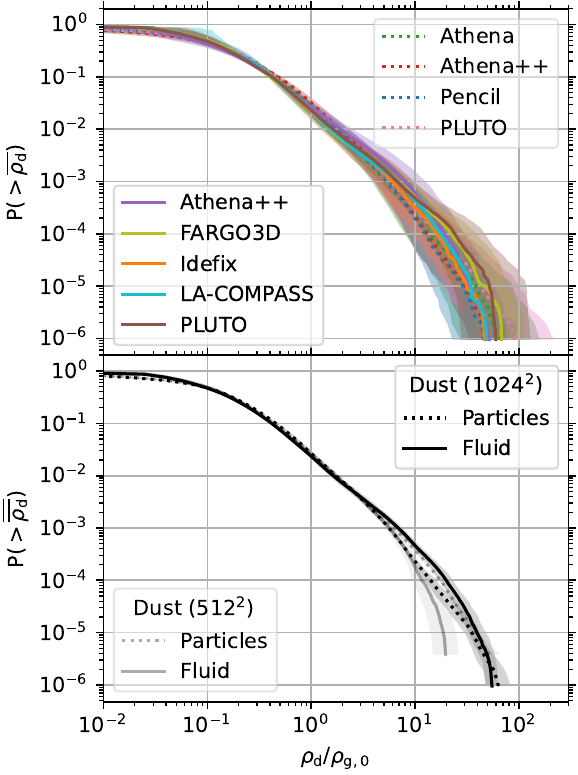}
    \caption{Similar to Figure~\ref{fig:BA-512_CDF}, except the upper panel only shows grid resolutions at $1024^2$, and the lower panel shows them in black along with $512^2$ in gray.}
    \label{fig:BA-1024_CDF}
\end{figure}

Figure~\ref{fig:BA-1024_CDF} is similar to Figure~\ref{fig:BA-512_CDF} for $512^2$.
With four times as many grid cells, $\Prob[\max(\rhod)]$ reaches $1024^{-2} \approx 1 \times 10^{-6}$ instead of $512^{-2} \approx 4\times10^{-6}$.
In the upper panel, $\Prob(>\avgrhod)$ diverges much less between codes overall at $1024^2$ than at $512^2$, including at $\rhod > 10\rhogn$.
In the lower panel, the dust density distributions at $1024^2$ resolution agree well between dust models, with $\max(\meanavgrhod)$ only about 13\% lower for fluids than for particles.
Compared to the mean distributions at $512^2$ included in the panel, this general agreement at $1024^2$ is predominantly due to the high density tail for dust fluids [i.e., $\Prob(>\meanavgrhod) < 10^{-3}$] shifting to the right with increased resolution, consistent with the resolution study of model~BA (Section~\ref{sec:problem_ba}) by \citet[][Fig. 10, lower left panel]{FARGO3D_fluid} for FARGO3D.
For all particle codes in the upper panel, the mean distribution in the lower panel shifts slightly to the left (i.e., toward lower densities) at $1024^2$ compared to $512^2$ for $\Prob(>\meanavgrhod) < 10^{-3}$, which was not found in a similar resolution study by \citet[][Fig.~6, lower left panel]{Athena_particles} for run~BA in Athena but with $\np = 9$.
However, we find convergence of the mean distribution for particles over all probabilities shared between resolutions [i.e., $\mathrm{P}(>\meanavgrhod) > 4\times10^{-6}$], consistent with the resolution study by \citet[][Fig.~A2, upper right panel]{BaronettYangZhu2024} for Athena++ with $\np = 4$.
We note that the significant time variability of the highest densities, as shown by the shaded areas in the upper panels and by $\max[\rhod(t)]$ in Figure~\ref{fig:BA-1024_time_series}, can make such comparisons less robust.

\begin{deluxetable*}{lllllrrrrrrrrrrrrr}
    \tablecaption{Execution and Performance\label{tab:performance}}
    \tablecolumns{17}
    \tablehead{
        \colhead{Code}      & \colhead{Dust}    & \colhead{User}    & \colhead{Microarchitecture}   & \colhead{$C$} & \multicolumn{4}{c}{$\Npe$}                                & \multicolumn{4}{c}{Core-Hours}                                & \multicolumn{4}{c}{Normalized Time Steps}                     \\
                            &                   &                   &                               &               & \multicolumn{2}{c}{$512^2$}   & & $1024^2$                & \multicolumn{2}{c}{$512^2$}   & & $1024^2$                    & \multicolumn{2}{c}{$512^2$}   & & $1024^2$                    \\
        \multicolumn{1}{c}{(1)}&\multicolumn{1}{c}{(2)}&\multicolumn{1}{c}{(3)}&\multicolumn{1}{c}{(4)}&\multicolumn{1}{c}{(5)}&\multicolumn{2}{c}{(6)}&&\multicolumn{1}{c}{(7)}& \multicolumn{2}{c}{(8)}       &&\multicolumn{1}{c}{(9)}       & \multicolumn{2}{c}{(10)}      &&\multicolumn{1}{c}{(11)}}
    \startdata
        Athena++            & Fluid$\cnote$     & P. Huang          & Intel Skylake                 & 0.30          & \multicolumn{2}{r}{64}        & \multicolumn{2}{r}{64}    & \multicolumn{2}{r}{75}        & \multicolumn{2}{r}{448}       & \multicolumn{2}{c}{179K}      & & \multicolumn{2}{r}{363K}    \\
        FARGO3D             & Fluid             & L. Krapp          & Intel Ivy Bridge              & 0.44          & \multicolumn{2}{r}{128}       & \multicolumn{2}{r}{128}   & \multicolumn{2}{r}{85}        & \multicolumn{2}{r}{486}       & \multicolumn{2}{c}{163K}      & & \multicolumn{2}{r}{326K}    \\
        Idefix$\anote$      & Fluid             & H. Aly            & Intel Golden Cove             & 0.80          & \multicolumn{2}{r}{256}       & \multicolumn{2}{r}{256}   & \multicolumn{2}{r}{50}        & \multicolumn{2}{r}{452}       & \multicolumn{2}{c}{347K}      & & \multicolumn{2}{r}{706K}    \\
                            &                   & G. Lesur          & Intel Skylake                 & 0.80          & \multicolumn{2}{r}{256}       & \multicolumn{2}{r}{512}   & \multicolumn{2}{r}{42}        & \multicolumn{2}{r}{409}       & \multicolumn{2}{c}{346K}      & & \multicolumn{2}{r}{695K}    \\
                            &                   &                   & NVIDIA Ampere$\enote$         & 0.80          & \multicolumn{2}{r}{6912}      & \multicolumn{2}{r}{6912}  & \multicolumn{2}{r}{2995}      & \multicolumn{2}{r}{19,930}    & \multicolumn{2}{c}{346K}      & & \multicolumn{2}{r}{695K}    \\
        LA-COMPASS$\bnote$  & Fluid             & S. Li             & AMD Zen 2                     & 0.80          & \multicolumn{2}{r}{1024}      & \multicolumn{2}{r}{4096}  & \multicolumn{2}{r}{512}       & \multicolumn{2}{r}{4915}      & \multicolumn{2}{c}{280K}      & & \multicolumn{2}{r}{557K}    \\
        PLUTO               & Fluid             & P. Sudarshan      & Intel Sunny Cove              & 0.30          & \multicolumn{2}{r}{128}       & \multicolumn{2}{r}{128}   & \multicolumn{2}{r}{30}        & \multicolumn{2}{r}{210}       & \multicolumn{2}{c}{172K}      & & \multicolumn{2}{r}{346K}    \\
        \hline
        &&&                                                                     &\multicolumn{1}{c}{$\np =$} &\multicolumn{1}{r}{1}&\multicolumn{1}{r}{9}&\multicolumn{2}{r}{1} &\multicolumn{1}{r}{1}&\multicolumn{1}{r}{9}&\multicolumn{2}{r}{1} &\multicolumn{1}{r}{1}&\multicolumn{1}{r}{9}&\multicolumn{2}{r}{1}\\
        \hline
        Athena              & Particles$\dnote$ & J. Lim            & Intel Skylake                 & 0.80          & 16    & 16                    & & 64                      & 27    & 100                   & & 336                         & 179K   & 178K                 & & 361K                        \\
                            &                   &                   &                               & 0.40          & 16    & \nodata               & & \nodata                 & 89    & \nodata               & & \nodata                     & 178K   & \nodata              & & \nodata                     \\
                            &                   &                   &                               & 0.40          & 256   & \nodata               & & \nodata                 & 300   & \nodata               & & \nodata                     & 178K   & \nodata              & & \nodata                     \\
        Athena++            & Particles         & S. A. Baronett    & Intel Skylake                 & 0.40          & 1     & \nodata               & & \nodata                 & 36    & \nodata               & & \nodata                     & 192K   & \nodata              & & \nodata                     \\
                            &                   &                   &                               & 0.40          & 16    & \nodata               & & 64                      & 54    & \nodata               & & 772                         & 185K   & \nodata              & & 364K                        \\
                            &                   &                   &                               & 0.40          & 256   & 256                   & & 1024                    & 276   & 1318                  & & 940                         & 182K   & 176K                 & & 357K                        \\
        Pencil              & Particles         & O. Brouillette    & Intel Golden Cove             & 1.00$\fnote$  & 1     & \nodata               & & \nodata                 & 52    & \nodata               & & \nodata                     & 288K   & \nodata              & & \nodata                     \\
                            &                   &                   &                               & 1.00          & 1     & \nodata               & & \nodata                 & 81    & \nodata               & & \nodata                     & 289K   & \nodata              & & \nodata                     \\
                            &                   &                   &                               & 1.00          & 16    & 64                    & & 64                      & 85    & 2674                  & & 1386                        & 292K   & 307K                 & & 551K                        \\
                            &                   &                   &                               & 0.50          & 16    & 64                    & & 64                      & 153   & 3877                  & & 2522                        & 274K   & 274K                 & & 550K                        \\
        PLUTO               & Particles         & M. Flock          & Intel Skylake                 & 0.45          & 8     & 8                     & & 16                      & 27    & 189                   & & 325                         & 173K   & 172K                 & & 348K 
    \enddata
    \tablecomments{Columns give the (1) code name, (2) dust model, (3) user, (4) processor microarchitecture, (5) Courant number $C$; total number of processing elements$\gnote$ $\Npe$ used at grid resolutions (6) $512^2$ and (7) $1024^2$; total core-hours$\hnote$ at (8) $512^2$ and (9) $1024^2$; and total number of integration time steps in thousands (K) at (10) $512^2$ and (11) $1024^2$, normalized to $C = 1.00$.
    Runs modeling the dust as a fluid or with particles are separated into upper and lower groups, respectively.
    The row separating these groups with horizontal rules specifies the average number of particles per grid cell/point $\np$ for the lower group, where Columns (6), (8), and (10) are split to differentiate runs at $512^2$ with either $\np = 1$ or 9.
    Additional rows for Idefix compare different microarchitectures, and those in the lower group compare different $C$ or domains decomposed across different $\Npe$ (Section~\ref{sec:performance}).\\
        $\anote$ For 2D problems, $C = 0.80$ in Idefix corresponds to $C = 0.40$ in PLUTO \citep[][\S~2.7 and eq.~12]{Idefix}.\\
        $\bnote$ Each run used 64 OpenMP threads (Footnote~\ref{foot:openmp}) in hybrid parallelization between MPI processes (Footnote~\ref{foot:mpi}).\\
        $\cnote$ Section~\ref{sec:pressureless_dust_fluid}\\
        $\dnote$ Section~\ref{sec:lagrangian_dust_particles}\\
        $\enote$ One NVIDIA A100 GPU has 6912 CUDA cores (Footnote~\ref{foot:cuda}).\\
        $\fnote$ With RK3; all other Pencil runs use RK4\\
        $\gnote$ Footnote~\ref{foot:pe}\\
        $\hnote$ Footnote~\ref{foot:core-hours}}
\end{deluxetable*}

\subsection{Performance} \label{sec:performance}

To help compare the computational performance of the various codes and numerical methods (Section~\ref{sec:codes_numerical_methods} and Table~\ref{tab:codes}), Table~\ref{tab:performance} summarizes the execution of the models presented in Sections~\ref{sec:fiducial_grid_resolution} and \ref{sec:higher_grid_resolution}.
Additional runs of Idefix (Section~\ref{sec:idefix}) were performed on different processor microarchitectures [Column~(4)], including a GPU.
Moreover, additional runs of some Lagrangian-dust codes (Section~\ref{sec:lagrangian_dust_particles}) were performed with different Courant numbers $C$ [Column~(5)] or with domains decomposed across different numbers of processing elements $\Npe$ [Columns~(6) and (7)].\footnote{
A processing element is any component (e.g., core or thread) of a microprocessor (e.g., CPU or GPU) that performs arithmetic and logic operations on data and connects to other processing elements through a network (-on-chip) or cache hierarchy \citep{Flynn1972, PattersonHennessyGoldberg1990}.\label{foot:pe}}

\begin{figure}
    \includegraphics[width=\columnwidth]{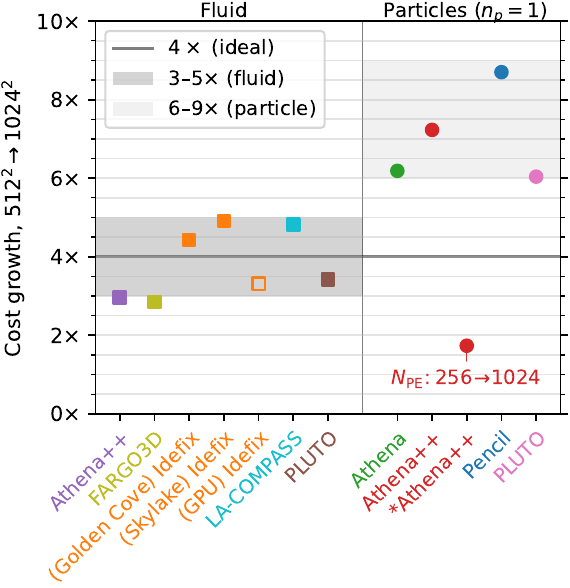}
    \caption{Growth in computational cost per normalized time step [core-hours in Columns~(8) and~(9) divided by normalized time steps in Columns~(10) and~(11) of Table~\ref{tab:performance}] from $512^2$ to $1024^2$ for each pair of runs, grouped by dust model.
    This ratio measures resolution scaling, not absolute speed; lower values do not indicate a faster code.
    Each ratio compares two runs of the same code on the same hardware with the same Courant number $C$, so differences in hardware, $\Npe$ (Footnote~\ref{foot:pe}), and $C$ between markers cancel.
    Circular or square markers show runs that implement $\np = 1$ particles or a dust fluid, respectively, while colors differentiate the codes; the open square marks the Idefix GPU run (cf. Figure~\ref{fig:performance_energy}).
    The solid gray line marks the ideal factor of four from the increase in total grid cells/points, while the shaded areas bracket the approximate ranges for fluid and particle codes, except for the labeled Athena++ particle runs whose $\Npe$ increases from $256$ to $1024$, as reported in Section~\ref{sec:performance}.}
    \label{fig:performance_resolution}
\end{figure}

To more fairly compare between runs using heterogeneous parallelization, we quantify execution times in Columns~(8) and (9) of Table~\ref{tab:performance} and in Figure~\ref{fig:performance_resolution} using core-hours \citep[e.g.,][]{Horwitz2024}.\footnote{
Multiplying the walltime $\TNpe$ (i.e., elapsed real time) by the number of processing elements used $\Npe$ (Footnote~\ref{foot:pe}) yields the total core-hours.\label{foot:core-hours}}
Although most of the runs used only MPI (Footnote~\ref{foot:mpi}) between processor cores, G. Lesur [Column~(3)] also ran Idefix on a single NVIDIA A100 GPU with 6912 CUDA cores (Footnote~\ref{foot:cuda}), and S. Li used hybrid parallelization to run LA-COMPASS (Section~\ref{sec:la-compass}) across 16 or 64 MPI processes, each with 64 OpenMP threads (Footnote~\ref{foot:openmp}).
Using core-hours ``flattens'' these differences to uniformly compare computational costs.

As expected and shown in Figure~\ref{fig:performance_resolution}, changes in grid resolution (Section~\ref{sec:higher_grid_resolution}) or particle resolution (Section~\ref{sec:higher_particle_resolution}) affect the computational cost.
Since $1024^2$ runs also take about twice as many time steps as $512^2$ runs (the time step halves with the cell size), we compare the cost per normalized time step [Columns~(10) and~(11)], for which ideal scaling with the four times more grid cells gives a factor of four.
For codes modeling a dust fluid [Column~(2)], runs at grid resolution $1024^2$ generally use about 3--5$\times$ more core-hours per normalized time step than at $512^2$.
For codes with an average of $\np = 1$ dust particle per grid cell (or point for Pencil; Section~\ref{sec:pencil}), $1024^2$ runs use about 6--9$\times$ more core-hours per normalized time step than at $512^2$, except for the Athena++ (Section~\ref{sec:athena++}) runs with $\Npe = 1024$ and 256, respectively, which used about 1.7$\times$ more.
The corresponding ratios of total core-hours are about twice these values.
Moreover, $512^2$ runs with $\np = 9$ used about 4--7$\times$ more core-hours than with $\np = 1$ but 25--32$\times$ more for Pencil.

The additional Idefix runs also contrast CPU and GPU scaling with resolution.
On both Intel CPU microarchitectures, Idefix used about 4.4--4.9$\times$ more core-hours per normalized time step at $1024^2$ than at $512^2$---near the upper end of the fluid range---whereas on a single NVIDIA A100 GPU it used only about 3.3$\times$ more ($\approx 30\%$ fewer), consistent with the more efficient scaling of GPUs over CPUs at higher resolutions \citep[][\S~5]{Idefix} and further reflected in the improved power efficiency (Section~\ref{sec:energy_efficiency-caveats}).

The Courant number also affects the total number of core-hours, as lowering $C$ shortens the simulation time difference $\Delta t$ between integration steps, thereby increasing the total number of time steps needed to reach the same $\tlim$.
For additional runs of Athena (Section~\ref{sec:athena}) and Pencil with the same $\Npe$, the total core-hours are inversely proportional to $C$.
Given this inverse scaling, we report the time steps normalized to $C = 1.00$ in Columns~(10) and (11) by multiplying the actual number of time steps by the actual value of $C$ used.
For two-dimensional problems such as BA (Section~\ref{sec:problem_ba}), we note that $C = 0.80$ in Idefix corresponds to $C = 0.40$ in PLUTO \citep[][\S~2.7 and eq.~12]{Idefix}.
Aside from Idefix, the number of normalized time steps for LA-COMPASS and Pencil exceeds all others by about 50\%.
Lastly, for the two serial ($\Npe = 1$) Pencil runs with $C = 1.00$, the one with RK3 time integration (Table~\ref{tab:performance}, Note~f) used about 36\% fewer core-hours and 500 fewer time steps than the one with RK4.

The imbalance of particle loads across processing elements, absent for pressureless fluids, increases synchronization overhead and can contribute to the significant difference in performance scaling (cf. Sections~\ref{sec:performance} and \ref{sec:scalability}) between dust models with respect to resolution (cf. Sections~\ref{sec:lagrangian_dust_particles} and \ref{sec:pressureless_dust_fluid}).
Although the total number of grid cells/points increases by four between $512^2$ and $1024^2$ for both dust models, the total number of particles also increases by four to keep $\np = 1$.
As the remaining dust filaments continue to consolidate throughout the saturated state from $20T$ to $100T$ (Figures~\ref{fig:BA-512_snapshots}, \ref{fig:BA-512-np9_snapshots}, and \ref{fig:BA-1024_snapshots}; Footnote~\ref{foot:videos}), an increasing number of processing elements assigned to regions with fewer and fewer particles must wait in real time until the next integration step for others to process an increasing majority of particles concentrated in progressively smaller regions.
This problem-specific (Section~\ref{sec:problem_ba}) load-balancing issue can also prolong the total core-hours for an increased number of particles at the same grid resolution (e.g., $\np = 9$ at $512^2$).

\subsubsection{Scalability} \label{sec:scalability}

In terms of scalability for high-performance computing, parallel efficiency can be defined as
\begin{equation}
    \pareff \equiv \frac{S}{\Npe},
    \label{eq:pareff}
\end{equation}
where $\Npe \in \mathbb{N}^+$ is the number of processing elements (e.g., cores or threads; Section~\ref{sec:performance}) used and speedup
\begin{equation}
    S(\Npe > 1) \equiv \frac{T_1}{\TNpe},
    \label{eq:S}
\end{equation}
in the context of Amdahl's law \citep{Amdahl1967}, for serial and parallel walltimes $T_1$ and $\TNpe$, respectively.
By substitution, we can redefine parallel efficiency as
\begin{equation}
    \pareff \equiv \frac{T_1}{\TNpe\Npe},
    \label{eq:pareffalt}
\end{equation}
where the denominator is simply the total core-hours for the parallelized execution (Footnote~\ref{foot:core-hours}).
In general, parallel walltimes scale as
\begin{equation}
    \TNpe \propto \Npe^{-\pareff}.
    \label{eq:walltime-scaling}
\end{equation}
Thus, for two parallelized runs, where $\Npe^\prime > \Npe$ with respective walltimes $T_{\Npe^\prime}$ and $\TNpe$, we can also define \textit{relative} efficiency (e.g., if $T_1$ is unknown) as
\begin{equation}
    \relpareff \equiv -\frac{\ln(T_{\Npe^\prime}/\TNpe)}{\ln(\Npe^\prime/\Npe)}.
    \label{eq:relpareff}
\end{equation}

\begin{figure}
    \includegraphics[width=\columnwidth]{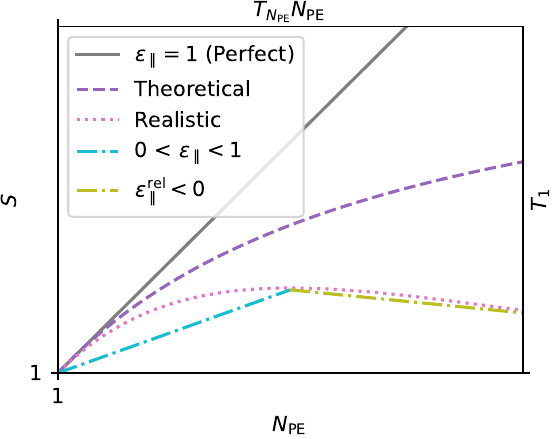}
    \caption{Schematic examples of parallel efficiency.
    The solid gray line shows perfect efficiency $\pareff = 1$ [equation~\eqref{eq:pareff}], i.e., speedup $S$ [equation~\eqref{eq:S}] as a linear function of the number of processing elements $\Npe$;
    the dashed purple curve theoretical (asymptotic) efficiency, according to Amdahl's law \citep{Amdahl1967};
    and the dotted pink curve realistic efficiency, accounting for coordination overhead (Section~\ref{sec:scalability}).
    The strictly positive slope of the cyan dashed--dotted line shows an example measurement of (absolute) $\pareff$ [equation~\eqref{eq:pareffalt}], e.g., using the serial walltime $T_1$ and the total core-hours $\TNpe\Npe$ (Footnote~\ref{foot:core-hours}) of a parallelized execution.
    The negative slope of the olive dashed--dotted line shows relative efficiency $\relpareff < 0$ [equation~\eqref{eq:relpareff}] beyond a critical $\Npe$.}
    \label{fig:pareff}
\end{figure}

Figure~\ref{fig:pareff} schematically shows the parallel efficiency (or strong scaling) of various cases for a fixed problem size.
As an example of perfect efficiency $\pareff = 1$ (i.e., linear $S$ or ideal strong scaling), parallelizing a serial task that takes $T_1 = 10$ real hours across $\Npe = 10$ would take $T_{10} = 1$ real hour or 10 core-hours.
However, unavoidable serial portions of the task will ultimately limit the potential efficiency of parallelization \citep{Amdahl1967}.
Thus, theoretically $S(\Npe\to\infty)$ is asymptotic, corresponding to diminishing returns in efficiency.
Moreover, parallelization in practice requires coordination overhead, e.g., associated with the scheduling of threads (e.g., OpenMP; Footnote~\ref{foot:openmp}) or communication between processing elements (e.g., MPI; Footnote~\ref{foot:mpi}).
Thus, $\relpareff$ can realistically become negative beyond a critical $\Npe$.

\begin{figure}
    \includegraphics[width=\columnwidth]{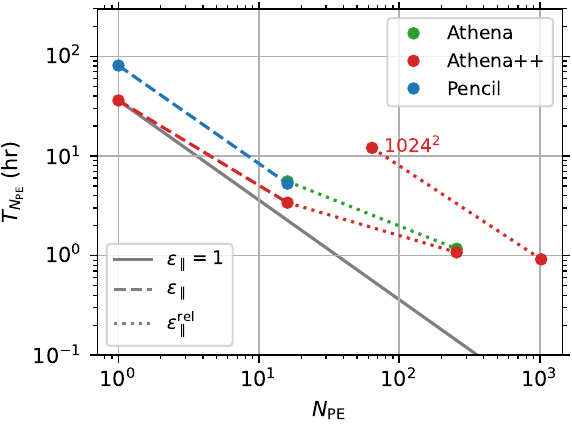}
    \caption{Measured strong scaling, i.e., walltime $\TNpe$ (Footnote~\ref{foot:core-hours}) vs. the number of processing elements $\Npe$ (Footnote~\ref{foot:pe}) for $\np = 1$ particle runs at $512^2$, and at $1024^2$ for the labeled Athena++ runs.
    As in Figure~\ref{fig:performance_resolution}, circular markers show runs that implement $\np = 1$ particles, while colors differentiate the codes.
    The solid gray line shows perfect efficiency $\pareff = 1$ [equations~\eqref{eq:pareff} and \eqref{eq:pareffalt}] from the serial Athena++ run, which parallel Athena++ runs at $512^2$ exceed by a factor of $1/\pareff$ [equation~\eqref{eq:walltime-scaling}].
    Dashed lines connect serial runs ($\Npe = 1$) to parallel runs of the same code, while dotted lines connect two parallel runs and have slopes of $-\relpareff$ [equation~\eqref{eq:relpareff}].
    The progressive flattening of the $512^2$ Athena++ curve toward larger $\Npe$ reflects the growing imbalance of particle loads discussed in Section~\ref{sec:scalability}.}
    \label{fig:performance_scaling}
\end{figure}

We can assess the parallel efficiency of some of the codes that model $\np = 1$ dust particles by comparing how computational times scale for different $\Npe$ in Table~\ref{tab:performance} and Figure~\ref{fig:performance_scaling}.
As an example of high efficiency, the Pencil runs with $C = 1.00$ and RK4 time integration (cf. Table~\ref{tab:performance}, Note f) at $512^2$ show $\pareff = 96\%$ for $\Npe = 16$, aided by improved load balancing through particle block domain decomposition (Section~\ref{sec:pencil}).
On the other hand, for the three separate runs of Athena++ at a resolution of $512^2$, $\pareff = 66\%$ for $\Npe = 16$ reduces to 13\% for $\Npe = 256$ due to increasingly imbalanced particle loads (Section~\ref{sec:performance}).
Although communication overhead also scales with $\Npe$, this specific loss of performance with $\Npe = 256$ results disproportionally from imbalanced loads, as the $\np = 1$ runs for Athena and Athena++ used at least five times more core-hours than the corresponding $512^2$ Idefix runs with $\Npe = 256$ and 25 times more for $\np = 9$ in Athena++ [Column~(8)].

Alternatively, we can assess the relative efficiency for some of the particle codes.
For Athena++, $\relpareff = 41\%$ between $\Npe = 16$ and 256.
However, at $1024^2$, $\relpareff = 93\%$ between $\Npe = 64$ and 1024 (also a 16-fold increase in $\Npe$), suggesting that the relative efficiency of Athena++ may scale better at higher resolutions (as an indirect way to better balance particle loads), as has also been observed for hydrodynamic codes that support GPUs (see below and Section~\ref{sec:energy_efficiency-caveats}).
Lastly, for Athena runs with $C = 0.40$ at $512^2$, $\relpareff = 56\%$ between $\Npe = 16$ and 256 is somewhat higher than that for Athena++ (between the same $\Npe$).

\subsubsection{Energy Efficiency and Caveats} \label{sec:energy_efficiency-caveats}

The pursuit of energy efficiency has influenced the shift in emphasis from CPUs to GPUs in the current era of ``exascale'' computing.\footnote{Exascale computing systems can calculate at least $10^{18}$ IEEE 754 \citep{IEEE2019} double-precision (64-bit) floating-point operations (additions or multiplications) per second (exaFLOPS).}
To demonstrate the substantial difference in efficiency, we can estimate and compare the energy consumed by Idefix runs in Table~\ref{tab:performance}.
As published by Intel, the maximum thermal design power (TDP)\footnote{
Thermal design power (TDP) is the expected amount of heat a processor can generate under sustained normal workloads that its cooling system must dissipate.
It can be used as a proxy to estimate the long-term average power rating of a processor \citep[][p.~22]{HennessyPattersonAsanovic2012}.}
of the Xeon Gold 6130 Skylake CPU [released in 2017; Column~(4)] with 16 cores per die is 125~W.
This amounts to 2 and 4~kW for the 256 and 512 CPU cores (16 and 32 chips or 8 and 16 dual-socket nodes) used at $512^2$ and $1024^2$ resolutions [Columns~(6) and (7)], respectively.
On the other hand, one standard Ampere A100 GPU (released in 2020) has a maximum TDP of 300~W, as published by NVIDIA.

\begin{figure}
    \includegraphics[width=\columnwidth]{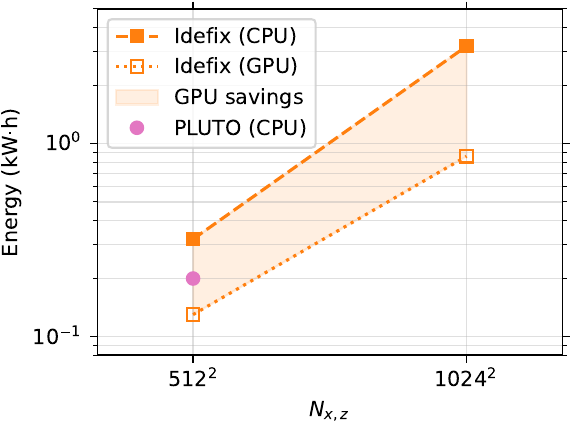}
    \caption{Estimated energy consumed in kilowatt-hours (Section~\ref{sec:energy_efficiency-caveats}) as a function of grid resolution $N_{x,z}$.
    As in Figures~\ref{fig:performance_resolution} and \ref{fig:performance_scaling}, circular or square markers show runs that implement $\np = 1$ particles or a dust fluid, respectively, while colors differentiate the codes.
    For Idefix, filled squares with dashed lines or open squares with dotted lines show runs on Intel Skylake CPUs or a single NVIDIA A100 GPU, respectively, while the shaded area estimates the energy saved by the GPU, which widens with resolution.
    The lowest-core-hour run in Table~\ref{tab:performance} (PLUTO, $\np = 1$, $512^2$) is included for reference.}
    \label{fig:performance_energy}
\end{figure}

To estimate the total energy used, as plotted in Figure~\ref{fig:performance_energy}, we multiply the total TDP by the walltimes of each execution.\footnote{
The walltime (i.e., the elapsed real time) can be recovered by dividing the total core-hours [Columns~(8) and (9)] by the number of processing elements used [Columns~(6) and (7), respectively; cf. Footnote~\ref{foot:core-hours}].}
At $512^2$ and $1024^2$, the walltimes for CPUs were 0.16 and 0.80 hours, respectively, whereas those for the GPU were 0.43 and 2.88 hours, respectively.
Thus, the estimated energy used for these $512^2$ and $1024^2$ runs was respectively 0.32 and $3.2~\kWh$ (kilowatt-hours) for CPUs but respectively only 0.13 and $0.86~\kWh$ for the GPU.
In other words, one GPU was at least two or three times more energy efficient than 256 or 512 CPU cores, respectively.

We can also compare against the run with the fewest core-hours overall at 26.8, i.e., PLUTO with $\np = 1$ dust particles at $512^2$ resolution on eight different Intel Skylake CPU cores.
With a TDP of 205~W for 28 cores on a single die (Xeon Platinum 8180, released in 2017), this run still used $0.20~
\kWh$ or at least 50\% more energy than the equivalent GPU run, as also shown in Figure~\ref{fig:performance_energy}.
Moreover, $512^2$ is toward the lower limit of resolution to warrant the use of GPUs over CPUs, as \citet[][\S~5]{Idefix} estimates efficiencies up to a factor of six in much higher resolution tests with $256^3$ subdomains for Idefix.
Although newer CPU architectures can be somewhat more energy efficient,\footnote{The AMD Zen 2 CPUs (EPYC 7H12, released in 2019) used for the LA-COMPASS runs use about 4.4~W per core, compared to 7.3~W per core for the Intel Skylake Xeon Platinum 8180.} we note that these estimates do not account for the additional energy consumed by the required cooling and networking systems across nodes, with CPUs typically requiring more nodes than GPUs.

As a caveat, the use of different processors (e.g., with different internal clock speeds), compilers (e.g., proprietary ones), as well as compiler options and optimizations (e.g., vectorization) can affect computational performance in this project and in general \citep[][\S~3.6.1]{Athena++}.
For example, among Idefix runs, the Intel Golden Cove microarchitecture was released in 2021, while Intel Skylake was released in 2015.
However, the use of the open-source GNU compiler on Golden Cove may have contributed to these runs taking about 10--20\% longer than those on Skylake using Intel's proprietary compiler.
Ideally, all codes compared here should use the same compiler with the same or no optimizations and be executed on the same hardware, which may be attempted in future code comparisons where feasible.
However, a more quantitative, let alone comprehensive, comparison of these differences is beyond the scope of this work.

\section{Discussion} \label{sec:discussion}

We discuss the broader implications of our results reported in Section~\ref{sec:results}.
Section~\ref{sec:stochasticity} explains that rapid divergence resulting from numerical differences between codes renders only statistical comparisons between simulations meaningful.
Section~\ref{sec:dust_models} examines the statistical behavior, numerical trade-offs, and physical implications of modeling dust as either Lagrangian particles or a pressureless fluid.
Section~\ref{sec:power_spectra} compares the power spectra of the dust density between codes, dust models, and resolutions, and quantifies the shot noise of Lagrangian particles.

\subsection{Stochasticity} \label{sec:stochasticity}

\begin{figure*}
    \centering
    \includegraphics[scale=1.044]{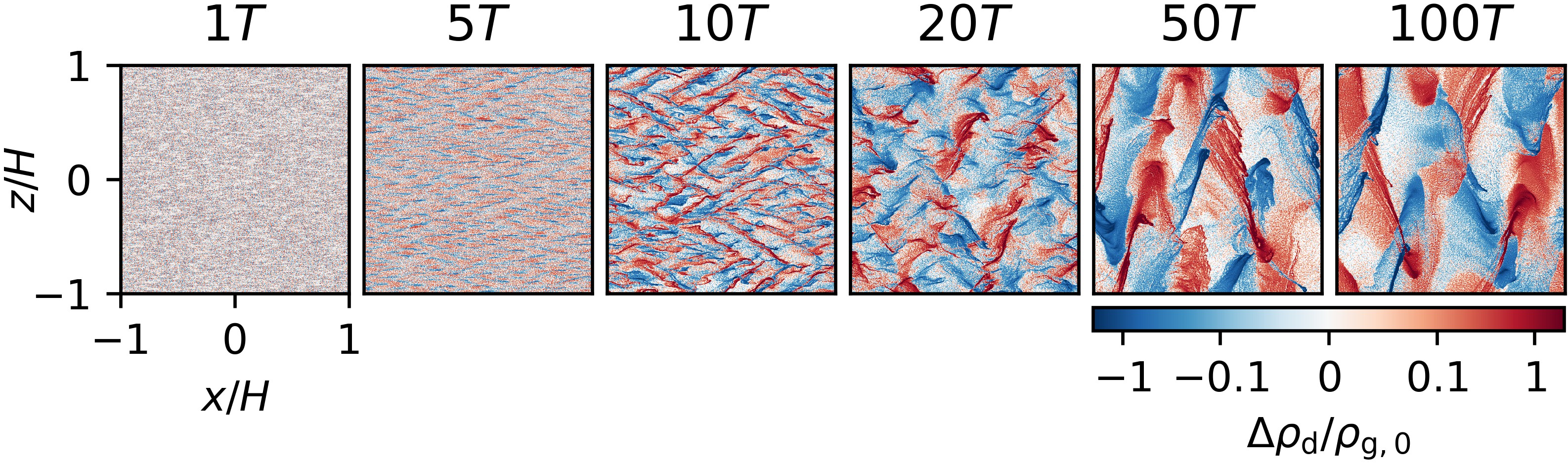}
    \caption{Similar to Figure~\ref{fig:BA-512_snapshots} except for the absolute difference in Lagrangian dust densities $\Delta\rhod$ between Athena++ and PLUTO models (i.e., the former minus the latter; cf. positive red values here with the upper Athena++ row in Figure~\ref{fig:BA-512_snapshots}), initialized with the same particle positions and velocities (at $t = 0$; Section~\ref{sec:stochasticity}), and with snapshots as early as $t = 1T$.}
    \label{fig:BA-512_snapshot_diffs}
\end{figure*}

To assess the simulation time at which identically initialized Lagrangian particle trajectories may diverge between codes, we ran an additional PLUTO model (Section~\ref{sec:pluto}) using the same particle positions and velocities at $t = 0$ in Athena++ (Section~\ref{sec:athena++}) with grid resolutions at $512^2$.
Figure~\ref{fig:BA-512_snapshot_diffs} visualizes their divergent evolution by mapping the absolute difference in dust density fields interpolated onto the grid (Section~\ref{sec:lagrangian_dust_particles}) over a series of snapshots.
Although the first snapshot appears noisy, we find relative differences as high as 200\% as early as $t = 1T$.

Although each model itself is deterministic (i.e., identical evolutions from the same initial state when compiled for and executed on the same hardware), this rapid divergence between codes is indicative of chaotic behavior due to the algorithms being different.
The streaming instability has long been known to lead to a turbulent saturated state \citep[][\S~3]{JohansenYoudin2007}.
Thus, the divergence of such dynamical systems is extremely sensitive to numerical differences in initial conditions or between integration steps, and matching the long-term behavior between codes is virtually impossible.

As shown by Table~\ref{tab:codes} and Section~\ref{sec:codes_numerical_methods}, the codes we feature use a variety of integration methods, which are not necessarily supported by all.
Differences in Riemann solvers, spatial reconstruction methods (including flux limiters), and time integrators can all contribute to rapid dynamical divergence (e.g., HLLE and Roe solvers or VL2 and RK2 integrators for Athena++ and PLUTO, respectively).
Even if these match, it is impractical to ensure that grid discretization (e.g., cell centers or edges), initial particle positions and velocities, or integration time steps $\Delta t$ are identical between codes to machine precision.
Moreover, the same portable code may even result in slightly different evolutions (e.g., Figure~\ref{fig:BA-512_snapshot_diffs}) when run on different processor microarchitectures or compiled with different optimizations (e.g., vectorization).
Therefore, the statistical comparisons shown in Section~\ref{sec:results} (e.g., cumulative distribution functions, time averaged over the saturated state) remain the most practical and robust for an inherently stochastic system.

\subsection{Dust Models} \label{sec:dust_models}

As detailed in Sections~\ref{sec:fiducial_grid_resolution} and \ref{sec:higher_grid_resolution}, Figures~\ref{fig:BA-512_time_series_pre-10T} and \ref{fig:BA-1024_time_series_pre-10T} show clear differences between the early evolution of dust modeled as Lagrangian particles (Section~\ref{sec:lagrangian_dust_particles}) or as a pressureless fluid (\ref{sec:pressureless_dust_fluid}).
Except for LA-COMPASS (\ref{sec:la-compass}), $\max[\rhod(t < 10T)]$ for fluids is about half that for particles but grows exponentially about $1T$ earlier.
For particles in Athena++ (Section~\ref{sec:athena++}), we offer no explanation for the $2T$ delayed onset of exponential growth compared to other particle codes, as also seen in Figure~\ref{fig:BA-512-np9_time_series_pre-10T}.
At fixed grid resolution, Figure~\ref{fig:BA-512-np9_time_series_pre-10T} also shows that higher-particle-resolution models converge (Section~\ref{sec:higher_particle_resolution}) toward the early evolution of dust fluids.

The differences in density distributions between dust models at the nonlinear saturated state may have implications for hydrodynamic studies of planetesimal formation.
The lower panel of Figure~\ref{fig:BA-512_CDF} shows that the time-averaged $\max(\meanavgrhod)$ reached by dust fluids is about 40\% of that reached by $\np = 1$ Lagrangian particles (Section~\ref{sec:problem_ba}), both at grid resolution $512^2$.
Moreover, the lower panel of Figure~\ref{fig:BA-512-np9_CDF} shows that increasing the particle resolution to $\np = 9$ appears to slightly lower the $\max(\meanavgrhod)$ of such models---consistent with a similar study by \citet[][Fig.~6, upper right panel]{Athena_particles}---but it remains nearly twice as high as that of fluids at the same grid resolution.
Thus, streaming instability simulations using dust fluids with similar parameters (e.g., $\taus = 1$; Section~\ref{sec:gas}) and effective grid resolutions of $256/H$ or lower (Table~\ref{tab:parameters}) might be less probable at meeting various planetesimal formation criteria, e.g., Hill densities \citep[][eq.~9]{LimSimonLi2025} or collapse under self-gravity (\citealt[][\S~2.1]{AbodSimonLi2019}; \citealt[][\S~3.2]{LimSimonLi2024}).
Furthermore, differences at the denser end of the cumulative distribution function---e.g., $\Prob(>\meanavgrhod)$ for $\rhod > 3\rhogn$ in Figure~\ref{fig:BA-512-np9_CDF}---may also affect the initial mass function of planetesimals \citep[][\S~3.1.3]{AbodSimonLi2019}.
Only at a higher grid resolution of $1024^2$ do dust fluids seem to be more in line with particles at the denser end, bringing their $\max(\meanavgrhod)$ to within about 15\% of that for particles (Figure~\ref{fig:BA-1024_CDF}).
However, it remains to be seen whether this remains consistent at even higher resolutions.
Overall, the already convergent particle density distributions between $512^2$ and $1024^2$ resolutions (Figures~\ref{fig:BA-512-np9_CDF} and \ref{fig:BA-1024_CDF})---consistent with similar studies by \citet[][Fig.~6, lower left panel]{Athena_particles} and \citet[][Fig.~A2, upper right panel]{BaronettYangZhu2024}---suggests that dust-fluid simulations should employ a minimum grid resolution to agree with particle solutions.

The convergence of the particle-density distributions at these resolutions also argues against artificial particle trapping.
\citet[\S~3.2.2]{PriceFederrath2010} found dense structures formed by Lagrangian particles used as passive tracers in the Eulerian code FLASH \citep{FryxellOlsonRicker2000, DubeyFisherGraziani2008}, which were absent from the grid-based density field, and time-averaged particle-density distributions that varied more strongly with resolution than the grid-density distributions \citep[\S~3.3.1]{PriceFederrath2010}.
\citet{GenelVogelsbergerNelson2013} showed that this bias is specific to massless tracers advected with the local gas velocity: because such tracers follow the reconstructed velocity field rather than the mass fluxes given by the Riemann solver, they accumulate wherever the reconstructed velocity is discontinuous across cell interfaces (their \S~3.2 and Figs.~7--8), and their density statistics do not converge with the number of tracers (their \S~4.1 and Fig.~10).
Nor is the interpolation order decisive, as \citet[\S~2.3.2]{PriceFederrath2010} found no strong dependence of their results on the tracer interpolation scheme \citep[see also][\S~2.1]{GenelVogelsbergerNelson2013}.
The dust particles here differ from such tracers in three respects.
First, they have inertia: with $\taus = 1$ (Section~\ref{sec:problem_ba}), each particle relaxes toward the local gas velocity over $\tstop = \Omega^{-1}$ [equations~\eqref{eq:par_acc} and \eqref{eq:par_acc_resid}], i.e., over hundreds of time steps (Table~\ref{tab:performance}), and so does not respond instantaneously to jumps in the interpolated gas velocity.
Second, they carry mass and exert drag back on the gas [equation~\eqref{eq:gas_mom}], so their density is part of the dynamics rather than a passive diagnostic.
Third, the streaming instability here drives subsonic turbulence without shocks \citep[][Fig.~8]{BaronettYangZhu2024}, unlike the supersonic turbulence\footnote{\citet{PriceFederrath2010} considered three-dimensional turbulence at an rms Mach number of 10 and grid resolutions of $128^3$, $256^3$, and $512^3$.} of \citet{PriceFederrath2010}.
Moreover, all particle codes here assign and interpolate with the TSC scheme (Sections~\ref{sec:athena}, \ref{sec:athena++}, \ref{sec:pencil}, and \ref{sec:pluto}), whose continuous, piecewise-quadratic kernel varies smoothly as a particle crosses cell boundaries, unlike the discontinuous nearest-grid-point assignment \citep[][Ch.~5]{HockneyEastwood1981}.
We therefore find no indication that artificial trapping contributes to the higher peak densities of the particle runs.

Various studies have investigated differences between modeling distributions of solids as Lagrangian particles or as pressureless fluids in other contexts---e.g., turbulent molecular clouds \citep{CommerconLebreuillyPrice2023}, pebble accretion \citep{ChrenkoChametlaMasset2024}, and debris disks \citep{LynchLovellSefilian2024}---yet important numerical considerations and physical implications remain.
Although Lagrangian particles avoid the discretization error inherent to Eulerian fluids arising from grid reconstruction (e.g., finite-volume flux inaccuracies), the interpolation required by particle--mesh methods to compute drag-induced particle trajectories or their collective back reaction introduces its own numerical error.
However, the second-order TSC kernel significantly improves over the first-order cloud-in-cell (CIC) scheme \citep{HockneyEastwood1981}.
Moreover, representing many particles as Lagrangian super-particles (Section~\ref{sec:lagrangian_dust_particles}) is itself a physical approximation that affects both the solid mass resolution (Section~\ref{sec:higher_particle_resolution}) and the parallel load-balancing performance (Section~\ref{sec:performance}).

On the other hand, approximating particles as a pressureless Eulerian fluid carries its own caveats.
In particular, it forgoes the underlying direction-dependent velocity dispersion and higher-order moments, while concentration and momentum diffusion terms could be added for an improved formulation \citep[e.g.,][]{WeberPerezBenitez-Llambay2019, Athena++_fluid}.
As in moment-based radiation-hydrodynamics methods---e.g., flux-limited diffusion \citep{LevermorePomraning1981} or the M1 closure \citep{Levermore1984}---where intersecting beams merge because radiation is treated as a fluid (\citealt[][Fig.~3]{WeihOlivaresRezzolla2020}; \citealt[][Fig.~2]{MelonFuksmanFlockKlahr2025}), the dust fluid approach likewise precludes crossing filaments or streams.
Such deficiencies may be related to the absence of resolution convergence in dust-fluid density distributions observed in prior streaming-instability studies where dust filaments collide throughout a field full of vortices \citep[e.g.,][Fig.~10, upper panels for model~AB]{FARGO3D_fluid}.
Furthermore, the extreme density gradients produced in our Pencil runs (Section~\ref{sec:pencil}) of Problem~BA (Section~\ref{sec:problem_ba}) with a dust fluid repeatedly caused the finite-difference code to crash before reaching $\tlim = 100T$; accordingly, we omit those results.
However, with the generalized mean field theory approach of \citet{Binkert2023} and a more recent development of a non-Newtonian dust-fluid model for astrophysical flows \citep{LynchLaibe2024}, more investigations in the numerical theory underlying either the particle-based or the fluid-based approach seem warranted.

\subsection{Power Spectra} \label{sec:power_spectra}
Given the stochasticity of this system (Section~\ref{sec:stochasticity}), statistical diagnostics provide the only meaningful basis for comparison between codes.
The cumulative distributions in Figures~\ref{fig:BA-512_CDF}, \ref{fig:BA-512-np9_CDF}, and \ref{fig:BA-1024_CDF} characterize how often each dust density occurs but not the spatial scales on which the densest structures form.
To resolve these scales, we compute the power spectrum of the dust density fluctuations $\delta\rhod \equiv \rhod - \langle\rhod\rangle$ [equation~\eqref{eq:vol_avg}], normalized such that its sum over all wavenumbers returns the variance $\langle(\delta\rhod)^2\rangle$ \citep[][\S~3.4.3]{PriceFederrath2010}.
Denoting the discrete Fourier transform by $\dhrhod \equiv \mathcal{F}(\delta\rhod)/(N_xN_z)$ \citep[cf.][eq.~13]{BaronettYangZhu2024}, we sum the power of all modes $\mathbf{k}^\prime$ within annuli of width $k_\mathrm{f} \equiv 2\pi/L_x = \pi/H$, the fundamental wavenumber of the domain (Section~\ref{sec:problem_ba}):
\begin{equation}
    E(k) \equiv \frac{1}{k_\mathrm{f}}\sum_{\left||\mathbf{k}^\prime| - k\right| < k_\mathrm{f}/2}\dhrhod(\mathbf{k}^\prime)\,\dhrhod^{\,*}(\mathbf{k}^\prime),
    \label{eq:power_spectrum}
\end{equation}
where $k = k_\mathrm{f}, 2k_\mathrm{f}, \ldots$ and the asterisk denotes the complex conjugate.
By Parseval's theorem, $\sum_k E(k)k_\mathrm{f} = \langle(\delta\rhod)^2\rangle$.
We omit the mean ($k = 0$) mode and plot annuli only up to the Nyquist wavenumber $\kNyq = \pi N_x/L_x$, i.e., $\kNyq H \approx 804$ and 1608 at $512^2$ and $1024^2$, respectively, although annuli beyond $\kNyq$, which include the corners of the two-dimensional Fourier domain, still count toward the variance.
For each run, we average $E(k)$ over the same saturated-state snapshots as the cumulative distributions to obtain $\avgE(k)$, again using the particle densities interpolated at grid centers for direct comparison with the dust fluids (Section~\ref{sec:fiducial_grid_resolution}).

Lagrangian particles sample the dust with a finite number of super-particles $N_\mathrm{p} = \np N_xN_z$ (Section~\ref{sec:lagrangian_dust_particles}), which adds Poisson (i.e., shot) noise to $\rhod$.
For equal-mass particles, this noise has an approximately white spectrum, filtered by the TSC assignment kernel (Sections~\ref{sec:athena}, \ref{sec:athena++}, \ref{sec:pencil}, and \ref{sec:pluto}), with an expected power of $\langle\rhod\rangle^2W^2(\mathbf{k})/N_\mathrm{p}$ per mode, where $W$ is the Fourier transform of the kernel.
This level depends on $\np$ and $\langle\rhod\rangle$ but not on the spatial distribution of the particles, so it applies throughout the evolution.
Rather than rely on this estimate, which neglects aliasing, we measure the shot-noise floor $\Eshot(k)$ directly from the initial state of each particle run.
Because the particle positions start uniformly random (Section~\ref{sec:problem_ba}), each initial ($t = 0$) density field consists of shot noise alone.
In every run, its variance agrees with the expected TSC-filtered Poisson value to within 1\%, and its spectrum exceeds the analytic TSC-filtered estimate by 36--52\% near $\kNyq$ because of aliasing.
We plot $\Eshot$ for comparison but do not subtract it from $\avgE$.

\begin{figure}
    \includegraphics[width=\columnwidth]{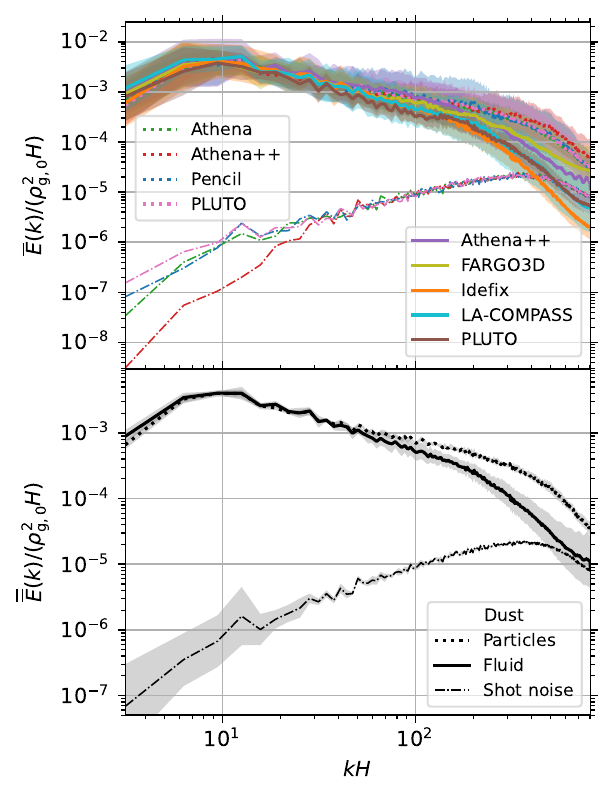}
    \caption{Power spectra $E(k)$ of the dust density fluctuations [equation~\eqref{eq:power_spectrum}] for Problem~BA at grid resolution $512^2$, with wavenumbers $k$ in units of the inverse gas scale height $H^{-1}$ and $E$ in units of $\rhogn^2H$.
    Dotted and solid lines show codes that implement an average of $\np = 1$ particle per grid cell/point and a pressureless dust fluid (Sections~\ref{sec:lagrangian_dust_particles} and \ref{sec:pressureless_dust_fluid}), respectively.
    In the upper panel, curves show time averages over the saturated state $\avgE$ (cf. Figure~\ref{fig:BA-512_CDF}), colors different codes (Section~\ref{sec:codes_numerical_methods}), shaded areas the $1\sigma$ time variability, and dash-dotted curves the shot-noise floor $\Eshot$ of each particle code, measured from its initial state (Section~\ref{sec:power_spectra}).
    In the lower panel, curves show the mean $\meanavgE$ of all particle or fluid codes in the upper panel, the dash-dotted curve the mean $\Eshot$, and shaded areas the standard deviation of their respective $\avgE$ or $\Eshot$.}
    \label{fig:BA-512_power_spectrum}
\end{figure}

Figure~\ref{fig:BA-512_power_spectrum} shows the spectra at our fiducial grid resolution of $512^2$.
In the upper panel, all codes peak near $kH \approx 10$ and agree within their time variability for $kH \lessapprox 50$, regardless of the dust model.
At smaller scales, the codes separate by dust model: in the lower panel, $\meanavgE$ for particles exceeds that for fluids at $kH \gtrapprox 50$, i.e., on wavelengths shorter than about 30 grid cells, by up to a factor of about three near $\kNyq$.
Within each dust model, the spread at high $k$ follows the high-density tails in Figure~\ref{fig:BA-512_CDF}, e.g., Idefix retains the least small-scale power and Athena++ with particles the most.
Every particle spectrum remains above its shot-noise floor at all $k$, with the smallest ratio $\avgE/\Eshot$ of 3.4--5.9 occurring just below $\kNyq$ ($kH \approx 750$--800).
Shot noise therefore contributes at most about 30\% of the particle power, even at the grid scale with $\np = 1$, too little to account for the gap between dust models.

\begin{figure}
    \includegraphics[width=\columnwidth]{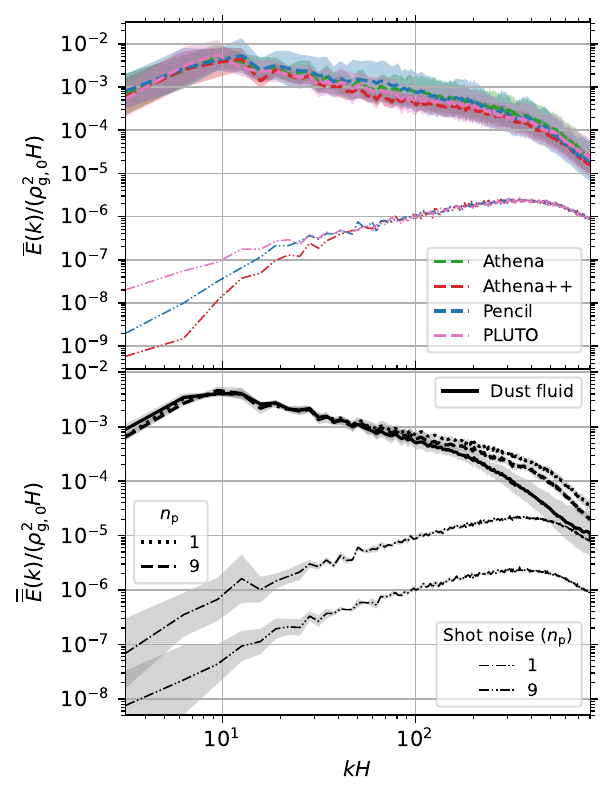}
    \caption{Similar to Figure~\ref{fig:BA-512_power_spectrum}, except the upper panel only shows particle codes with $\np = 9$, with dash-dot-dotted curves for their shot-noise floors $\Eshot$, and the lower panel shows their mean $\meanavgE$ with a dashed black curve, along with the means for $\np = 1$ (dotted) and dust fluids (solid) and the mean $\Eshot$ for $\np = 1$ (dash-dotted) and 9 (dash-dot-dotted).
    Shot-noise floors are shown only for particle runs with an initial-state snapshot.}
    \label{fig:BA-512-np9_power_spectrum}
\end{figure}

Figure~\ref{fig:BA-512-np9_power_spectrum} tests this conclusion with $\np = 9$.
Nine times as many particles lower $\Eshot$ by the same factor, so the smallest ratio $\avgE/\Eshot$ rises to 17--24, and shot noise contributes at most about 6\% of the particle power.
In the lower panel, $\meanavgE$ for $\np = 9$ closely follows that for $\np = 1$ at $kH \lessapprox 50$ and falls below it toward $\kNyq$, partly because of the lower shot noise, yet it remains above the dust-fluid mean at $kH \gtrapprox 50$.
Hence, the excess small-scale power of the particles at $512^2$ does not arise from shot noise.
After subtracting each floor near $\kNyq$, the remaining power agrees between $\np = 1$ and 9 to within about 15\% for Pencil and PLUTO, whereas Athena++ retains about 2.7 times more with $\np = 1$, a difference we cannot yet explain.

\begin{figure}
    \includegraphics[width=\columnwidth]{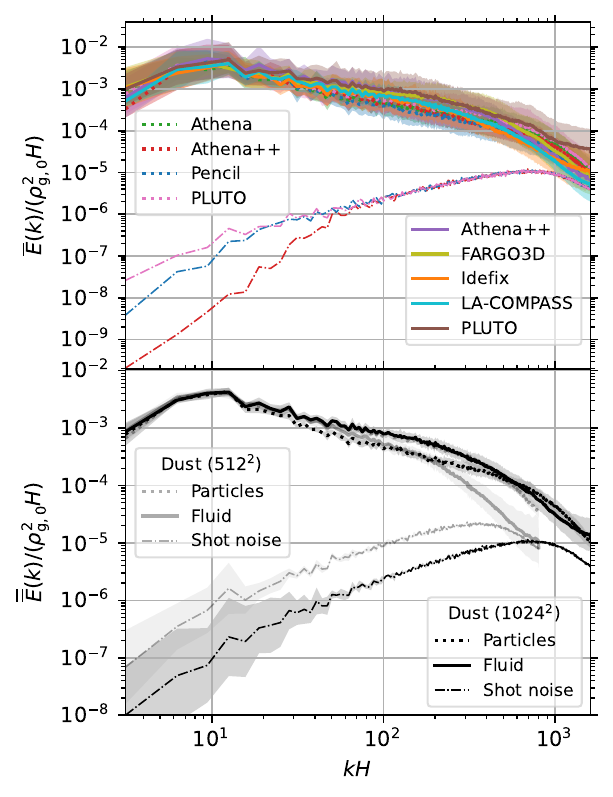}
    \caption{Similar to Figure~\ref{fig:BA-512_power_spectrum}, except the upper panel only shows grid resolutions at $1024^2$, and the lower panel shows them in black along with $512^2$ in gray.
    Shot-noise floors are shown only for particle runs with an initial-state snapshot.}
    \label{fig:BA-1024_power_spectrum}
\end{figure}

Figure~\ref{fig:BA-1024_power_spectrum} compares the spectra at $1024^2$ with those at $512^2$.
Doubling the resolution extends the dust-fluid spectra to smaller scales: in the lower panel, $\meanavgE$ for fluids at $1024^2$ rises above that at $512^2$ for $kH \gtrapprox 30$ and exceeds the particle mean at intermediate scales before converging with it near $\kNyq$, whereas the particle means at both resolutions nearly coincide where they overlap.
This behavior parallels Figure~\ref{fig:BA-1024_CDF}, where the high-density tail of the dust fluids shifts toward higher densities with resolution while that of the particles appears converged (Section~\ref{sec:higher_grid_resolution}).
With $\np = 1$ fixed, quadrupling the number of particles lowers $\Eshot$ by a factor of four at low $k$, but the finer grid now reaches wavenumbers where $\avgE$ is lower still.
Consequently, the smallest ratio $\avgE/\Eshot$ drops to 1.7--3.2 near $\kNyq$ ($kH \approx 1600$), and shot noise contributes up to about 60\% of the particle power at the grid scale, so we interpret the particle spectra there with caution.

Together, these spectra show on which scales the dust models differ (Section~\ref{sec:dust_models}).
On large scales ($kH \lessapprox 30$), all codes agree at both resolutions regardless of the dust model, whereas at $512^2$ the dust fluids lack power on the scales of the thinnest filaments, consistent with their lower peak densities (Figure~\ref{fig:BA-512_CDF}).
Because this deficit persists against the low-noise $\np = 9$ spectra and narrows with grid resolution, we attribute it to the dust fluids rather than to the particles, e.g., to numerical diffusion near the grid scale or to the inability of a pressureless fluid to represent crossing streams (Section~\ref{sec:dust_models}).
We do not fit power-law slopes, since none of the spectra follows a single power law between its peak and the dissipation or noise scale, and a fitted slope would largely reflect the choice of fitting range \citep[cf.][\S~3.4.3]{PriceFederrath2010}.

\section{Conclusions} \label{sec:conclusions}

This work presents the first systematic, multi-code comparison of the nonlinear saturation of the streaming instability.
In this comparison, we use an unstratified shearing box with a dimensionless stopping time of unity.
By analyzing seven hydrodynamic codes---spanning finite-volume and finite-difference schemes and implementing dust as either Lagrangian particles or a pressureless fluid---we have quantified both the common features and the resulting differences that arise from methodological choices.
Despite substantial differences in numerical algorithms, reconstruction schemes, and drag-coupling strategies, all codes reproduce the characteristic sequence of exponential growth, filament formation, and turbulent saturation.
This agreement lends credence to the nonlinear simulations of the streaming instability presented in the literature thus far.

At the same time, our results demonstrate that the dust model remains the dominant source of variation at moderate resolution.
At $512^2$, particle-based simulations systematically reach higher peak densities and exhibit broader high-density tails in their cumulative distribution functions than fluid-based models.
Increasing the particle resolution (e.g., by a factor of nine) reduces Poisson noise and brings the early evolution into closer alignment with dust-fluid runs, but the saturated-state density distributions remain distinct.
Only at higher grid resolution ($1024^2$) do the two dust treatments converge toward similar high-density statistics, with differences reduced to about 15\%.
These findings imply that dust-fluid simulations of the streaming instability may require higher spatial resolution than particle-based simulations to achieve comparable fidelity in the nonlinear regime.

Our performance analysis further reveals that computational cost scales differently across codes and dust models.
Particle-based simulations without block domain decomposition incur additional overhead from load imbalance as dust concentrates into progressively smaller regions, whereas fluid-based models scale more strongly with resolution.
GPU-accelerated implementations, such as Idefix on NVIDIA A100 hardware, demonstrate significantly improved energy efficiency and competitive scaling at high resolution, underscoring the growing importance of heterogeneous architectures for large-scale dust--gas simulations.

Finally, we emphasize that the stochasticity of this system limits the usefulness of trajectory-level comparisons between codes.
Even when initialized with identical particle positions and velocities, solutions diverge within a single orbital period.
Consequently, statistical measures (e.g., time-averaged density distributions) provide the only meaningful basis for cross-code validation.
Moreover, power spectra of the dust density show that the dust fluids lack small-scale power at $512^2$, a deficit that shot noise cannot explain and that closes at $1024^2$.
Nonetheless, the broad agreement we find in these statistical diagnostics, particularly at higher resolution, suggests that the streaming instability is numerically well captured across a diverse set of modern hydrodynamic frameworks.

Future work should extend this comparison to stratified domains and regimes with shorter stopping times, where dust-fluid approaches are expected to perform more favorably.
The present comparison establishes a foundation for such efforts and provides a benchmark for continued development of dust--gas simulation methods.

\begin{acknowledgments}
We thank Zhaohuan Zhu and Anna C. Childs for helpful discussions and suggestions that improved this work.
SAB acknowledges funding from the UNLV Foundation Board of Trustees Fellowship.
SAB used Microsoft 365 Copilot (Claude model, Anthropic; September 2026) to assist in drafting portions of the revised text, including Sections~\ref{sec:dust_models} and \ref{sec:power_spectra}; all content was reviewed, verified, and approved by the authors.
WL, JBS, OMU, CCY, and ANY acknowledge support from NASA award 80NSSC21K0497.
HA and SJP acknowledge funding from the European Research Council (ERC) under the European Union’s Horizon 2020 research and innovation programme (grant agreement No 101054502).
LE acknowledges support from NASA award 80NSSC25K7117.
PH acknowledges the National Science Foundation of China under grant Nos. 12503070, 12233004 and 12533011.
JL, DGR, and JBS acknowledge support from the Emerging Worlds program through award 80NSSC20K0702.
JL also acknowledges support from the Future Investigators in NASA Earth and Space Science and Technology program through award 80NSSC22K1322.
DGR also acknowledges support from the Future Investigators in NASA Earth and Space Science and Technology program through award 80NSSC24K1833.
PS acknowledges the support of the German Science Foundation (DFG) through grant number 495235860 and is a Fellow of the International Max Planck Research School for Astronomy and Cosmic Physics at the University of Heidelberg (IMPRS-HD).
CCY also acknowledges support from NASA through the Emerging Worlds program (award 80NSSC23K0653) and the Astrophysics Theory Program (80NSSC24K0133).
Resources supporting this work were provided by the NASA Advanced Supercomputing (NAS) Division at Ames Research Center.
We acknowledge computational support from the Max Planck Computing and Data Facility.
\end{acknowledgments}

\begin{contribution}


SAB wrote and revised the manuscript; ran simulations; managed and coordinated project planning and execution; collected, investigated, analyzed, and validated the data; produced the figures; and curated and maintains the public repositories related to the project.
WL, JBS, CCY, OU, and ANY proposed, acquired funding for, and supervised the project and guided the evolving objectives.
WL, JBS, CCY, JL, RL, DC, and S-JP critically reviewed the manuscript and provided helpful suggestions for its development and improvement.
OB drafted Section~\ref{sec:pencil} and ran simulations.
MF and PS ran simulations and PS drafted Section~\ref{sec:pluto}.
PH drafted the paragraph on the implementation of a pressureless dust fluid in Section~\ref{sec:athena++} and ran simulations.
LK drafted Section~\ref{sec:fargo3d} and ran simulations.
GL drafted Section~\ref{sec:idefix} and ran simulations.
SL drafted Section~\ref{sec:la-compass} and ran simulations.
JL drafted Section~\ref{sec:athena} and ran simulations.
All coauthors reviewed and commented on the manuscript.
\end{contribution}

\software{
Athena \citep{Athena}, Athena++ \citep{Athena++}, FARGO3D \citep{FARGO3D}, Idefix \citep{Idefix}, LA-COMPASS \citep{LA-COMPASS}, Matplotlib \citep{Matplotlib}, Microsoft 365 Copilot \citep{Copilot2026}, NumPy \citep{NumPy}, Pandas \citep{PandasSoftware, PandasPaper}, Pencil \citep{Pencil}, PLUTO \citep{PLUTO}
          }


\appendix

\section{Public Repositories} \label{appx:public_repository}

To make this comparison as accessible as possible, we have created a public GitHub repository\footnote{ \url{https://github.com/pfitsplus/sicc}}
\citep{GitHubRepo} to maintain the relevant resources.
These include the source and input files needed to run the publicly available codes and dust models presented in Section~\ref{sec:results}, as well as Jupyter Notebooks to generate all of the figures.
Note that files may be added after this publication as they become publicly available and as we explore future comparisons.
We invite users to create a GitHub issue in the repository for any questions related to these items, problems encountered in their use, or to provide feedback.

URLs for the repositories of public codes and dust modules featured here are listed in Table~\ref{tab:repos}.
However, we note that their maintainability or the validity of the URLs cannot be guaranteed beyond the time of this publication.

\begin{deluxetable*}{lll}
    \tablecaption{Public Repositories\label{tab:repos}}
    \tablecolumns{3}
    \tablehead{
        \colhead{Code}  & \colhead{Dust Module} & \colhead{URL} \\
        \multicolumn{1}{c}{(1)}&\multicolumn{1}{c}{(2)}&\multicolumn{1}{c}{(3)}}
    \startdata
        Athena          & Included              & \url{https://github.com/PrincetonUniversity/Athena-Cversion}    \\
        Athena++        & \nodata               & \url{https://github.com/PrincetonUniversity/athena}             \\
                        & Fluid                 & \url{https://github.com/PinghuiHuang/athena-multifluid-dust}    \\
        FARGO3D         & Included              & \url{https://github.com/FARGO3D/fargo3d}                        \\
        Idefix          & Included              & \url{https://github.com/idefix-code/idefix}                     \\
        Pencil          & Included              & \url{https://github.com/pencil-code/pencil-code}                \\
        PLUTO           & \nodata               & \url{https://plutocode.ph.unito.it}
    \enddata
    \tablecomments{Columns give (1) the code name, (2) if the dust module is included or available separately, and (3) the URL of the corresponding repository.}
\end{deluxetable*}

\bibliography{refs}{}
\bibliographystyle{aasjournalv7}



\end{document}